\documentclass[epj]{svjour}
\usepackage{graphicx}
\usepackage{overpic}
\usepackage[intlimits]{amsmath}
\usepackage{amssymb}
\usepackage{slashed}
\usepackage{cite}
\newcommand{\beq}{\begin{equation}}
\newcommand{\eeq}{\end{equation}}
\newcommand{\beqa}{\begin{eqnarray}}
\newcommand{\eeqa}{\end{eqnarray}}
\newcommand{\no}{\nonumber}

\newcommand{\newop}[2]{\def#1{\mathop{\mathrm{#2}}\nolimits}}
\newop{\artanh}{artanh}
\newop{\diag}{diag}
\newop{\Re}{Re}
\newop{\Im}{Im}
\begin{document}
\title{A gauge invariant chiral
unitary framework for kaon photo- and electroproduction on the proton}
\author{B.~Borasoy\inst{1}, P.~C.~Bruns\inst{1}, U.-G.~Mei{\ss}ner\inst{1,2} \and R.~Ni{\ss}ler\inst{1} %
}                     
\institute{Helmholtz-Institut f\"ur Strahlen- und Kernphysik (Theorie)
Universit\"at Bonn, Nu{\ss}allee 14-16, D-53115 Bonn, Germany 
 \and 
 Institut f\"ur Kernphysik (Theorie), Forschungszentrum J\"ulich
D-52425 J\"ulich, Germany}
\date{Received: date / Revised version: date}
%
\abstract{
We present a gauge invariant approach to photoproduction of mesons on nucleons
within a chiral unitary framework. The interaction kernel for meson-baryon
scattering is derived from the chiral effective Lagrangian and
iterated in a Bethe-Salpeter equation. Within the leading order approximation
to the interaction kernel, data on kaon photoproduction from SAPHIR, CLAS
and CBELSA/TAPS are analyzed in the threshold region. The importance of 
gauge invariance and  the precision of various approximations in the 
interaction kernel  utilized in earlier works are discussed. 
} 
%

\authorrunning{B.~Borasoy {\it et al.}}
\titlerunning{A gauge invariant chiral
unitary framework for kaon photo- and electroproduction on the proton}
\maketitle
%
\section{Introduction} \label{intro}

The hadronic spectrum is still the least understood property
of QCD.  Most theoretical models predict much more states 
than are actually observed  so far in experiments, see {\it e.g.}\  
\cite{Capstick:1986bm,Glozman:1997ag,Loring:2001kv}. This
is sometimes called the `missing resonance problem'.
The search for missing resonances has therefore been an important goal
of various experimental efforts. The properties of baryon resonances are presently
under thorough investigation at several facilities, {\it e.g.}, at ELSA, JLab, MAMI
GRAAL, COSY and SPring-8. Due to their hadronic decay modes, however, 
many baryon resonances have large overlapping
widths which makes it a difficult task to study individual states.
In this respect, polarization observables can be used as a tool
to filter out specific resonances in specific reactions.

A possible explanation of the missing resonances could be that these states do not
couple strongly to the pion-nucleon channels which have provided to a large extent 
the resonance data. A strong coupling of these resonances to channels with
strange particles could be unraveled in photoproduction processes of
$K\Lambda$ and $K\Sigma$. Such experiments have been recently undertaken
at SAPHIR \cite{Glander:2003jw,Lawall:2005np} and CLAS \cite{McNabb:2003nf,Bradford:2005pt} 
with high precision. More recently, $K^0 \Sigma^+$ has been measured with the 
CB/TAPS detector at ELSA \cite{Castelijns:2007qt}.
These data indicate that resonances so far undetected might have been observed,
but an unambiguous theoretical interpretation is still lacking. In addition,
the beam polarization asymmetry for $\gamma p \to K^+ \Lambda, K^+ \Sigma^0$ 
has been investigated by the LEPS collaboration at SPring-8~\cite{Zegers:2003ux}.

Based on an analysis
of cross sections, beam asymmetries, and recoil polarizations, the Bonn-Gatchina
resonance model, {\it e.g.}, demands among others the $P_{11}(1840)$, $D_{13}(1870)$, 
$D_{13}(2170)$ resonances \cite{Anisovich:2005tf,Sarantsev:2005tg}, 
whereas in the model of \cite{Usov:2005wy} a $P_{13}(1830)$ is preferred.

Electroproduction experiments are even more sensitive to the structure
of the nucleon due to the longitudinal coupling of the virtual photon to the
nucleon spin and might in addition yield some insight into the possible
onset of perturbative QCD, see {\it e.g.}\ \cite{Carlson:1988gt,Leinweber:1992hy}. But
at low photon virtuality experimental data for $\gamma p \to K \Lambda, K \Sigma$
are still not available.

Apparently, there are sufficient data to be met by theoretical approaches.
In addition to the  photoproduction data,
further experimental constraints are provided by pion-induced reactions on the proton.
Any theoretical approach that aims to describe the photoproduction data, ought to be
consistent with the corresponding pion-in\-duced data. 

A successful theoretical approach to  meson-baryon scattering is provided
by chiral unitary methods, see {\it e.g.}\  
\cite{Kaiser:1995eg,Kaiser:1995cy,Oset:1997it,Oller:2000fj,Lutz:2001yb,Oller:2000ma}. 
In this framework the chiral effective Lagrangian
is utilized to derive, for example, the interaction kernel in a Bethe-Salpeter equation (BSE)
which iterates meson-baryon re{\-}scattering to infinite order. The BSE generates
resonances dynamically, hence, without their explicit inclusion 
the importance of resonances can be studied. Chiral unitary
approaches have been implemented quite successfully for photoproduction
processes, see {\it e.g.}\ refs.~ \cite{Kaiser:1996js,Nacher:1998mi,Caro_Ramon:1999jf,Borasoy:2002mt}, 
but as a simplification only those diagrams
were taken into account where the photon is absorbed first and then the
produced meson-baryon pair undergoes final state interaction. This simplified
treatment violates, in general, gauge invariance. In order to guarantee
gauge invariance, diagrams with the photon coupling to any intermediate state of the
meson-baryon bubble chain must be taken into account \cite{Borasoy:2005zg,Kvinikhidze:1998xn}. One goal
of the present work is to study the importance of these additional contributions which render the
amplitude gauge invariant, but have been neglected in previous works.

Another simplification in chiral unitary approaches is the reduction of the interaction kernel
to the on-shell point. Although the interaction kernel appears in loops,
it has been argued that the off-shell components can be absorbed by 
redefining the coupling constants \cite{Oller:1997ti}. 
In this work, we do not employ the on-shell approximation but  
present two alternative methods to retain the off-shell components
in the interaction kernel. For a discussion of these issues, we refer
the reader to ref.~\cite{Nieves:1999bx}.  

In chiral unitary approaches to meson photoproduction, it has been common
practice to utilize $s$-wave projections for the meson-baryon scattering
kernel and the photoproduction multipoles. We will also analyze the 
accuracy of this approximation since in our approach some higher partial
waves are generated through the small components of the Dirac spinors
describing the baryon octet fields and through the kaon (baryon) pole term in 
charged  (neutral) meson production.

The main goal of the present study is the construction of a minimal approach
to meson photoproduction based on the chiral effective Lagrangian
which is exactly unitary and gauge invariant. The presented method
fulfills these important requirements from field theory, while at the same time any
subset of diagrams cannot be omitted as this would violate unitarity or gauge invariance.
In this study, we restrict ourselves to the chiral effective Lagrangian
at leading order. The inclusion of higher chiral orders in the interaction kernel
is straightforward and necessary to obtain better agreement with experiment,
particularly at higher energies away from the respective thresholds.
This will be the subject of forthcoming work.

The manuscript is organized as follows. In the next section, the effective Lagrangian
and the Bethe-Salpeter formalism including off-shell components are introduced.
The gauge invariant extension to photo- and electroproduction processes is 
discussed in sect.~\ref{sec:photo}. Section~\ref{sec:results} contains the comparison
with experimental data on kaon photoproduction. The phenomenological impact
of the violation of gauge invariance and the use of the on-shell approximation 
is also discussed. We summarize our findings in sect.~\ref{sec:conclusions},
while lengthy formulae and a second, alternative method for including off-shell pieces
in the interaction kernel are relegated to the appendices.

\section{Bethe-Salpeter equation}
\label{sec:bse}
\def\theequation{\arabic{section}.\arabic{equation}}
\setcounter{equation}{0}

The chiral effective Lagrangian incorporates symmetries and symmetry-breaking
patterns of QCD in a model-inde{\-}pendent way, in particular chiral symmetry
and its explicit breaking through the finite quark masses. By expanding Green's
functions in powers of Goldstone boson masses and small momenta a chiral
counting scheme can be established. 
However, the strict perturbative chiral expansion is only applicable
at low energies, and it certainly fails in the vicinity of
resonances. In this respect, the combination of the chiral effective Lagrangian
with non-perturbative schemes based on coupled channels and the Bethe-Salpeter
equation (BSE) have proven useful both in the purely mesonic and in the
meson-baryon sector 
\cite{Kaiser:1995eg,Kaiser:1995cy,Oset:1997it,Oller:2000fj,Lutz:2001yb,Oller:2000ma}. 
Such approaches extend the range of 
applicability of the chiral effective Lagrangian by implementing exact
two-body unitarity in a non-perturbative fashion and generating  
resonances dynamically.

In this work we restrict ourselves to the meson-baryon Lagrangian at leading order
\beqa \label{eq:L}
\mathcal{L}_{\phi B}^{(1)} &=& \langle \bar B([i\slashed{D},B]
- m_{0}B) \rangle  \nonumber\\
&-&\frac{D}{2}\langle \bar B\gamma^{\mu}\gamma_{5}\lbrace u_{\mu},B\rbrace \rangle 
 - \frac{F}{2}\langle \bar B\gamma^{\mu}\gamma_{5}[u_{\mu},B]\rangle \,,
\eeqa
where the matrix $B$ collects the ground state baryon octet, $\langle\ldots\rangle$ 
denotes the trace in flavor space, $m_{0}$ is the common baryon octet mass in the 
chiral limit and 
\beq
[D^{\mu},B] = \partial^{\mu}B+[\Gamma^{\mu},B]
\eeq
is the covariant derivative of the baryon field with the `chiral connection'
\beq
\Gamma^{\mu} = \frac{1}{2}[u^{\dagger},\partial^{\mu}u]
-\frac{i}{2}\left( u^{\dagger}v^{\mu}u+uv^{\mu}u^{\dagger}\right) \,. 
\eeq
The  external vector field is given by $v_{\mu} = -eQ\mathcal{A}_{\mu}$ where 
$Q = \frac{1}{3}\diag(2,-1,-1)$ is the quark charge matrix.
The lowest lying octet of pseudoscalar meson fields $\phi$ enters the Lagrangian, 
in matrix form, as
\beq \label{eq:Uphi}
U(\phi) = u^{2}(\phi) = \exp(\sqrt{2}i\frac{\phi}{f_{0}}) \,,
\eeq
where $f_{0}$ denotes the pseudoscalar decay constant in the chiral limit. 
Furthermore, we use
\begin{eqnarray*}
u_{\mu}       &=& iu^{\dagger}\nabla_{\mu}Uu^{\dagger} \,,\\ 
\nabla_{\mu}U &=& \partial_{\mu}U-i[v_{\mu},U] \,,
\end{eqnarray*}
and the coupling constants $D=0.8, F=0.46$ \cite{DFvalues}.
By expanding the chiral connection in powers of the meson fields, one derives
from the effective Lagrangian the leading order $\phi^{2}\bar B B$ vertex (the
so-called `Weinberg-To\-mo\-za\-wa' (WT) term), which we use as the driving term in our 
Bethe-Salpeter equation. Stated differently, this vertex insertion is the 
`interaction kernel' of the integral equation. One finds for the corresponding 
potential $V$ (which is the WT-vertex graph multiplied by $i$)
\beq
V^{bj,ai}(\slashed{q}_{2},\slashed{q}_{1}) = g^{bj,ai} (\slashed{q}_{1} + \slashed{q}_{2}) \,.
\eeq
Here, $q_{1}$ and $q_{2}$ are the four-momenta of the incoming and the outgoing meson, respectively, 
and the coupling constants are summarized as a matrix in channel space, with the entries
\beq \label{eq:WTcoupling}
g^{bj,ai} = -\frac{1}{4f_{0}^{2}}
\langle \lambda^{b\dagger}[[\lambda^{j\dagger},\lambda^{i}],\lambda^{a}]\rangle \,,
\eeq
where  $\lambda^{a}$ are the generators of the $SU(3)$ Lie-Algebra 
in the physical (particle) basis. In this representation,
a double index $bj$ specifies a particular channel consisting of a baryon $b$ and a meson $j$. 
The baryon and the meson propagator, $iS$ and $i\Delta$, are also summarized as (diagonal) 
matrices in channel space. They are given by
\beqa
iS^{bj,ai}(\slashed{p}) &=& \frac{i\delta^{ba}\delta^{ji}}{\slashed{p}-m_{a}} \,,\\
i\Delta^{bj,ai}(p)      &=& \frac{i\delta^{ba}\delta^{ji}}{p^{2}-M_{j}^{2}}   \,.
\eeqa
We can now write down the integral equation for the meson-baryon scattering amplitude 
$T^{bj,ai}$ in a rather compact form (suppressing the channel indices, but remembering 
the matrix character of the various amplitudes):
\begin{multline} \label{eq:BSE}
T(\slashed{q}_{2},\slashed{q}_{1};p) = V(\slashed{q}_{2},\slashed{q}_{1}) \\
+  \int\frac{d^{d}l}{(2\pi)^{d}}V(\slashed{q}_{2},\slashed{l})iS(\slashed{p}-\slashed{l})\Delta(l)
   T(\slashed{l},\slashed{q}_{1};p) \,.
\end{multline}
\begin{figure}[t]
\includegraphics[width=9cm]{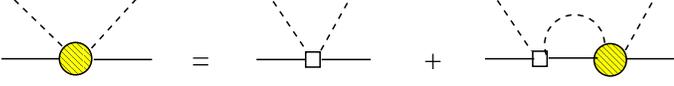}
\caption{Graphical illustration of the BSE for meson-baryon scattering. 
The filled circle represents the full scattering matrix and the open square
the driving meson-baryon vertex.}
\label{fig:BSE}
\end{figure}%
Here, $p\equiv p_{1}+q_{1}=p_{2}+q_{2}$ is the overall momentum, where $p_{1}$ and $p_{2}$ 
are the four momenta of the incoming and outgoing baryon, respectively. The BSE is illustrated 
in fig.~\ref{fig:BSE}. 
Note that we use dimensional regularization throughout. The solution of the BSE reads
\begin{multline} \label{eq:BSEsol}
T(\slashed{q}_{2},\slashed{q}_{1};p)= W(\slashed{q}_{2},\slashed{q}_{1};p) \\
+ W(\slashed{q}_{2},\tilde{p};p)G(p)[1-W(\tilde{p},\tilde{p};p)G(p)]^{-1}W(\tilde{p},\slashed{q}_{1};p)
\end{multline}
with
\begin{multline} \label{eq:W}
W(\slashed{q}_{2},\slashed{q}_{1};p) 
= \slashed{q}_{2}g\frac{1}{1+I_{\mathrm{M}}g} +\frac{1}{1+gI_{\mathrm{M}}}g\slashed{q}_{1} \\
-g \frac{1}{1+I_{\mathrm{M}}g}I_{\mathrm{M}}(\slashed{p}-m)\frac{1}{1+gI_{\mathrm{M}}}g \,.
\end{multline}
$W(\slashed{q}_{2},\tilde{p};p)$ can be obtained from eq.~(\ref{eq:W}) by replacing 
$\slashed{q}_{1} \rightarrow \slashed{p}-m \equiv \tilde{p}$, and so on, 
$m$ is the baryon mass matrix with entries
\beq
 m^{bj,ai} = \delta^{ba}\delta^{ji}m_{a} \,,
\eeq
and the loop integrals are given by
\begin{eqnarray}
\label{eq:IM} I_{\mathrm{M}}^{bj,ai} &=& \int\frac{d^{d}l}{(2\pi)^{d}}i\Delta^{bj,ai}(l) \,,  \\ 
\label{eq:G}  G^{bj,ai}(p)  &=& \int\frac{d^{d}l}{(2\pi)^{d}}i[\Delta(l)S(\slashed{p}-\slashed{l})]^{bj,ai} \,.
\end{eqnarray}
Note that in eqs.~(\ref{eq:BSEsol}) and (\ref{eq:W}), the symbol `1' represents the unit matrix
in channel space, while matrix-valued denominators denote matrix inversion.
In order to verify that the above expression for $T$ indeed solves the BSE, one employs the relation
\begin{multline}
\int\frac{d^{d}l}{(2\pi)^{d}}V(\slashed{q}_{2},\slashed{l})iS(\slashed{p}-\slashed{l})\Delta(l)
  W(\slashed{l},\slashed{q};p) \\
= V(\slashed{q}_{2},\tilde{p})G(p)W(\tilde{p},\slashed{q};p)+W(\slashed{q}_{2},\slashed{q};p)
  -V(\slashed{q}_{2},\slashed{q})
\end{multline}
for arbitrary $q$. Substituting the expression for $T$ in the BSE, eq.~(\ref{eq:BSE}) and 
using the last relation, it is straightforward to confirm that $T$ solves the integral equation, 
as claimed. It is important to keep in mind that all the involved expressions are matrices in 
channel space, so {\it e.g.}\  $I_{\mathrm{M}}$ does in general not commute with the coupling matrix $g$. 
The on-shell substitutions
\beq
\slashed{q}_{1,2}\rightarrow \slashed{p}_{1,2}+\slashed{q}_{1,2}-m = \slashed{p}-m \equiv \tilde{p} \,,
\eeq
which occur in the above expressions, provide the connection to the alternative method of 
solving the BSE presented in App.~\ref{app:offshell}. The solution given there is completely 
equivalent to the one described above. Here, again, we note that $\tilde{p}$ is a matrix in 
channel space due to the matrix character of the baryon mass matrix $m$.

Putting the external momenta on their respective mass shells, the form of the solution $T$ 
of eq.~(\ref{eq:BSE}) simplifies to
\beq \label{eq:Ton}
T_{\textrm{on}} = [1-WG]^{-1}W \,,
\eeq
where we used the shorthand notation $W\equiv W(\tilde{p},\tilde{p};p)$ and $G\equiv G(p)$. 
To obtain the complete on-shell meson-baryon scattering amplitude, this expression has to 
be sandwiched between baryon spinors $\bar u(p_{2})\ldots u(p_{1})$. 
Neglecting the tadpole integrals $I_{\mathrm{M}}$ in $W$, 
one obtains from eq.~(\ref{eq:Ton})
\beq
T'_{\textrm{on}} = [1-VG]^{-1}V \,,
\eeq
which is the form for the scattering amplitude usually encountered when utilizing the 
`on-shell-approximation' (which consists of setting also internal loop momenta in the 
potential $V$ on-shell) to arrive at a simplified version of the BSE. The integral equation 
eq.~(\ref{eq:BSE}) then reduces to an algebraic equation that can be solved by matrix inversion 
to give an expression like $T'_{\textrm{on}}$ above. Retaining the off-shell pieces in the exact 
solution $T_{\textrm{on}}$ therefore corresponds to keeping the terms proportional to $I_{\mathrm{M}}$ 
in the expression for $W$. 

In \cite{Oller:1997ti} it was argued that the on-shell
reduction of the interaction kernel is justified as the off-shell pieces
can be absorbed into the vertices. While this procedure simplifies
the interaction kernel and the treatment of the BSE, it is in fact
not necessary in order to solve the BSE as explained above.
We choose not to discard the off-shell pieces and evaluate the diagrams
in the way dictated by field theory, keeping in mind, 
however, that the off-shell parts are not uniquely fixed by any physical
requirements and depend on the field parametrization \cite{Nieves:1999bx}.
Recall that eq.~(\ref{eq:Uphi}) is not the only
possible choice for the parametrization of the meson fields, and that the
off-shell amplitudes in ChPT depend on this choice.

Another important reason for taking the full off-shell dependence of the interaction 
kernel into account is that the implementation of the BSE in 
a gauge-invariant description of photoproduction processes
is then straightforward. This is because we evaluate the occurring loop graphs 
according to the rules of field theory which also ensures that gauge symmetry can
be incorporated in a standard way.
This is not the case when utilizing the on-shell approximation.

At present, our interaction kernel is restricted to the lowest order contact term
derived from the effective meson-baryon Lagrangian. 
The extension to more general vertex structures in the kernel is possible
and will be discussed in future work. For now, it is not our aim to achieve perfect
agreement with experimental data (in which case higher-order terms in the kernel
would be indispensable), but to construct the simplest possible amplitude for kaon 
photoproduction that reconciles the framework of chiral unitary approaches with the 
fundamental principle of gauge invariance in a straightforward way.

For the sake of completeness, we add some comments on unitarity here. From the 
expression for $T_{\textrm{on}}$, eq.~(\ref{eq:Ton}), one immediately confirms the condition
for two-particle unitarity
\beq
\Im(T_{\textrm{on}}^{-1})= -\Im(G) \,.
\eeq
Our approach to photoproduction presented in the next section will be such that
unitarity is also satisfied when the photon is coupled to the meson-baryon bubble chain summed up
by means of the BSE, see also  \cite{Borasoy:2005zg}. 

The form of the solution $T_{\textrm{on}}$ guarantees that the above unitarity statement
holds. However, the careful reader might have noticed that the integrals $I_{\mathrm{M}}$ and
$G(p)$ occurring in the amplitude are divergent for $d\rightarrow 4$
and that the introduction of appropriate counterterms might spoil the simple form
of the solution given in eq.~(\ref{eq:Ton}). To deal with this complication one can
show by means of a rather lengthy calculation that the 
modification 
\beq
V\rightarrow V+\delta V\equiv V_{\delta} \,,
\eeq
of the interaction kernel V in the BSE, where 
\begin{align}
&V_{\delta}(\slashed{q}_{2},\slashed{q}_{1};p)= W_{\delta}(\slashed{q}_{2},\slashed{q}_{1};p) \no \\
&- W_{\delta}(\slashed{q}_{2},\tilde{p};p)\delta G(p)[1+W_{\delta}(\tilde{p},\tilde{p};p)\delta G(p)]^{-1}
   W_{\delta}(\tilde{p},\slashed{q}_{1};p) 
\end{align}
and
\begin{align}
W_{\delta}(\slashed{q}_{2},\slashed{q}_{1};p) &= \slashed{q}_{2}g\frac{1}{1-\delta I_{\mathrm{M}}g}
 +\frac{1}{1-g\delta I_{\mathrm{M}}}g\slashed{q}_{1} \no \\
&\quad+ g\frac{1}{1-\delta I_{\mathrm{M}}g}\delta I_{\mathrm{M}}(\slashed{p}-m)\frac{1}{1-g\delta I_{\mathrm{M}}}g  
\end{align}
leads to a new integral equation with a solution $T_{\delta}$.
The solution $T_{\delta}$ is obtained from $T$ by replacing
\begin{eqnarray*}
G     &\rightarrow & G-\delta G  \\
I_{\mathrm{M}} &\rightarrow & I_{\mathrm{M}}-\delta I_{\mathrm{M}}  
\end{eqnarray*}
in eqs.~(\ref{eq:BSEsol}) and (\ref{eq:W}). One observes that $V_{\delta}$ has the same 
form as $T$, with $G$ and $I_{\mathrm{M}}$ replaced by $-\delta G$ and $-\delta I_{\mathrm{M}}$,
respectively. The corrections $\delta G$ and $\delta I_{\mathrm{M}}$ have the form of
polynomial terms which serve to absorb the divergences in the loop
integrals. The divergences have thus been shifted from the loop integrals in
$T$ to the new kernel. Of course, this is not a
renormalization scheme in the usual sense, since the additional terms $\delta V$ in $V_{\delta}$
do not correspond exactly to counterterms derived from an effective
Lagrangian which is obvious form the lack of crossing symmetry of the amplitude. It
has already been noted in ref.~\cite{Nieves:1999bx}
that the solution of the BSE can not be renormalized within a usual renormalization scheme. However, the 
foregoing discussion shows at least that altering the loop integrals appearing in the 
solution of the BSE by terms $\delta G$ and $\delta I_{\mathrm{M}}$ is equivalent to certain 
modifications of the potential. In practice, we merely omit the divergences in the 
loop integrals in our expression for $T$, which will therefore depend on the regularization 
scale $\mu$ showing up in the modified loop integrals. 

This form of the off-shell BSE solution $T$ is not particularly convenient for the purpose 
of implementing it into the photoproduction amplitude. Therefore, we rewrite it by using the 
following decompositions:
\beqa
\label{eq:G01} G(p) &=& G_{1}(p)\slashed{p}+G_{0}(p) \,, \\
W(\tilde{p},\tilde{p};p) &=& W_{1}(p)\slashed{p} + W_{0}(p) \,, 
\eeqa
where the matrices $G_{0,1}$ follow straightforwardly from the explicit expression 
for the loop integral $G$ which is given in eq.~(\ref{eq:Gexplizit}) of 
App.~\ref{app:loop_integrals}, and
\beqa
W_{1}(p) &=& g\frac{2+I_{\mathrm{M}}g}{[1+I_{\mathrm{M}}g]^{2}} \,,  \\
W_{0}(p) &=& g\frac{1}{1+I_{\mathrm{M}}g}(I_{\mathrm{M}}m)\frac{1}{1+gI_{\mathrm{M}}}g \no \\
& &  -mg\frac{1}{1+I_{\mathrm{M}}g}-\frac{1}{1+gI_{\mathrm{M}}}gm \,. 
\eeqa
Note that $G_{i}(p)$ and $W_{i}(p)$ are Lorentz scalars which depend only on the variable $p^{2}$. 
Using these expressions one can derive
\beq
1-W(\tilde{p},\tilde{p};p)G(p) = \tilde W_{1}\slashed{p}+\tilde W_{0} \,,
\eeq
with
\beqa
\tilde W_{1} &=& -(W_{1}(p)G_{0}(p)+W_{0}(p)G_{1}(p)) \,,\\
\tilde W_{0} &=& 1-p^{2}W_{1}(p)G_{1}(p)-W_{0}(p)G_{0}(p) \,.
\eeqa
The above results yield
\beqa
G(p)[1-W(\tilde{p},\tilde{p};p)G(p)]^{-1} = \Omega_{1}(p)\slashed{p}+\Omega_{0}(p) 
\eeqa
with
\begin{align}
\Omega_{1}(p) &= G_{0}(p)[p^{2}\tilde W_{1}-\tilde W_{0}\tilde W_{1}^{-1}\tilde W_{0}]^{-1} \no \\
&\quad-G_{1}(p)\tilde W_{1}^{-1}\tilde W_{0}[p^{2}\tilde W_{1}-\tilde W_{0}\tilde W_{1}^{-1}\tilde W_{0}]^{-1} \,,\\
\Omega_{0}(p) &= p^{2}G_{1}(p)[p^{2}\tilde W_{1}-\tilde W_{0}\tilde W_{1}^{-1}\tilde W_{0}]^{-1} \no \\
&\quad-G_{0}(p)\tilde W_{1}^{-1}\tilde W_{0}[p^{2}\tilde W_{1}-\tilde W_{0}\tilde W_{1}^{-1}\tilde W_{0}]^{-1} \,.
\end{align}
All these results can be combined to provide the decomposition of the off-shell BSE-solution 
into the following independent Clifford algebra structures:
\beqa \label{eq:Tdecomp}
T(\slashed{q}_{2},\slashed{q}_{1};p) &=& \slashed{q}_{2}\slashed{p}\slashed{q}_{1}T_{1}(p) 
+ \slashed{q}_{2}\slashed{q}_{1}T_{2}(p) + \slashed{p}\slashed{q}_{1}T_{3}(p) \no \\
&& + \slashed{q}_{2}\slashed{p}T_{4}(p) + \slashed{q}_{1}T_{5}(p)+ \slashed{q}_{2}T_{6}(p)\no\\
&& + \slashed{p}T_{7}(p)+T_{8}(p) \,.
\eeqa
The scalar coefficient functions $T_{n}(p)$ read
\beq \label{eq:T_i}
\begin{split}
T_{1}(p) &= L_{1}\Omega_{1}(p)L_{1} \,, \\
T_{2}(p) &= L_{1}\Omega_{0}(p)L_{1} \,, \\
T_{3}(p) &= [L_{2}\Omega_{0}(p)+L_{3}\Omega_{1}(p)]L_{1} \,, \\
T_{4}(p) &= T_{3}^{T}(p) \,, \\
T_{5}(p) &= [p^{2}L_{2}\Omega_{1}(p)+L_{3}\Omega_{0}(p)]L_{1}+L_{1} \,, \\
T_{6}(p) &= T_{5}^{T}(p) \,, \\
T_{7}(p) &= [p^{2}L_{2}\Omega_{1}(p)+L_{3}\Omega_{0}(p)]L_{2} \\
&\quad+ [L_{2}\Omega_{0}(p)+L_{3}\Omega_{1}(p)]L_{3}^{T} -gI_{\mathrm{M}}L_{2} \,, \\
T_{8}(p) &= p^{2}[L_{2}\Omega_{0}(p)L_{2}+L_{3}\Omega_{1}(p)L_{2}+L_{2}\Omega_{1}(p)L_{3}^{T}] \\
&\quad+ L_{3}\Omega_{0}(p)L_{3}^{T}-L_{3}I_{\mathrm{M}}g \, ,
\end{split}
\eeq
where the superscript $T$ denotes transposition of channel indices and
\begin{eqnarray*}
L_{1} &=& g\frac{1}{1+I_{\mathrm{M}}g} \,, \\
L_{2} &=& \frac{1}{[1+gI_{\mathrm{M}}]^{2}}g \,, \\
L_{3} &=& -\frac{1}{1+gI_{\mathrm{M}}}gm\frac{1}{1+I_{\mathrm{M}}g} \,.
\end{eqnarray*}
Note that $L_{1}^{T}=L_{1}$ and $L_{2}^{T}=L_{2}$, but in general $L_{3}^{T}\not= L_{3}$.
For external on-shell particles, we can make the usual substitutions
\begin{displaymath}
\slashed{q}_{1},\slashed{q}_{2} \rightarrow \slashed{p}-m \equiv \tilde{p}
\end{displaymath}
to arrive at the on-shell scattering amplitude
\beq 
T_{\textrm{on}} = T_{\textrm{on}}^{(1)}\slashed{p}+T_{\textrm{on}}^{(0)}
\eeq
with
\beqa
T_{\textrm{on}}^{(1)} &=& p^{2}T_{1}+mT_{1}m-mT_{2}-T_{2}m \no \\
&& - T_{3}m-mT_{4}+T_{5}+T_{6}+T_{7} \,, \\
T_{\textrm{on}}^{(0)} &=& p^{2}(T_{2}+T_{3}+T_{4}-T_{1}m-mT_{1})+mT_{2}m \no \\
&& - T_{5}m-mT_{6}+T_{8} \,. 
\eeqa
The functions $T_{n}$ given in eq.~(\ref{eq:T_i}) depend only on the variable $p^{2}$.
They will enter the calculation of the various photo- and electroproduction amplitudes 
in the next section.

\section{Photo- and electroproduction}
\label{sec:photo}
\def\theequation{\arabic{section}.\arabic{equation}}
\setcounter{equation}{0}

The Bethe-Salpeter approach discussed in the last section 
(or the Lippmann-Schwinger equation in the non-relativistic
framework) can be implemented in 
electroproduction processes of mesons on the nucleon. In previous work, the
electromagnetic meson production on the nucleon was calculated at tree level
and the produced meson-baryon pair was subject to final-state interactions 
\cite{Kaiser:1995cy,Bennhold:1996th,Borasoy:2002mt}. As the photon does not couple to
all intermediate states of the meson-baryon bubble chain, gauge invariance
is in general violated and must be restored via artificial manipulations.
 Therefore it seems desirable to develop a formalism which implements the principles 
of gauge invariance and unitarity in a most natural and straightforward manner.
We follow here the path which we have already outlined in \cite{Borasoy:2005zg},
in rather general terms, for the case of a photon coupling to a meson-baryon 
scattering amplitude. In this work, we shall be more explicit in evaluating the 
contributions to the various amplitudes in question.
Our approach for constructing a unitary and gauge invariant electroproduction amplitude
decomposes into two major steps:
\begin{itemize}
\item[(1)] Fix the hadronic part of the amplitude by making use of a BSE to implement exact 
two-body unitarity.
\item[(2)] Couple the photon to the `hadronic skeleton' construc{\-}ted in step (1) wherever possible,
 {\it i.e.}\  to all external and internal lines 
describing the propagation of the involved particles as well as to (momentum-dependent) vertices.
\end{itemize}
The procedure of step (2), which is the most natural way to guarantee gauge invariance 
of the amplitude, leads to contributions that were usually not considered in chiral unitary 
approaches involving electromagnetic interactions. The importance of these additional
contributions which render the electroproduction amplitude gauge invariant
can also be quantified within the scheme utilized here.
The coupling of the photon to internal lines in the bubble chain generated by the BSE
leads to rather involved expressions, since the meson-baryon scattering amplitude $T$
appears twice in the corresponding amplitudes, as will be made more explicit
later when we give the formal expressions for the various contributions.

Exact unitarity (in the subspace of meson-baryon states)
is a fundamental principle satisfied by our approach. It is important to note that the procedure 
of step (2) above does {\em not} spoil the requirement of unitarity we built in by means of the BSE 
in step (1), as we will show below. 
First, we have to  specify  the set of graphs which constitute our amplitude.

\begin{figure}[t]
\includegraphics[width=9cm]{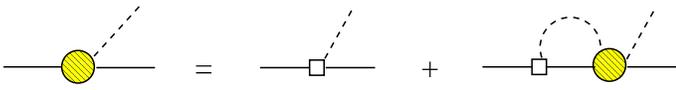}
\caption{The dressed meson-baryon vertex in the turtle approximation. The filled
         circle (open square) denotes the full (tree level) meson-baryon interaction.}
\label{fig:Gamma}
\end{figure}

We start with the tree level $\bar B\phi B$ amplitude derived from the leading order Lagrangian 
of eq.~(\ref{eq:L}),
\beq
\hat V^{bi,a} = \slashed{q}\, \gamma_{5}\, \hat g^{bi,a} \,.
\eeq
The last index $a$ labels the incoming baryon (here always the proton), while the double index 
 belongs to the outgoing pair of baryon $b$ and meson $i$. For fixed $a$,
$\hat V$ is a vector in channel space. Furthermore, $q$ is the four-momentum of the outgoing meson and
\beq
\hat g^{bi,a} = -\frac{D}{\sqrt{2}f_{0}}\langle \lambda^{b\dagger}\{\lambda^{i\dagger},\lambda^{a}\}\rangle 
- \frac{F}{\sqrt{2}f_{0}}\langle \lambda^{b\dagger}[\lambda^{i\dagger},\lambda^{a}]\rangle \,.
\eeq
To this tree level amplitude, we add the loop contribution that accounts for the final-state interaction 
after the meson has left the vertex, to obtain ({\it cf.}\ fig.~\ref{fig:Gamma})
\beq \label{eq:Gamma}
\Gamma(\slashed{q},\slashed{p}) =  \slashed{q}\gamma_{5}\hat g 
+ \int\frac{d^{d}l}{(2\pi)^{d}}T(\slashed{q},\slashed{l};p)iS(\slashed{p}-\slashed{l})\Delta(l)
  \slashed{l}\gamma_{5}\hat g \,.
\eeq
\noindent We call this approximation to the full meson-baryon interaction the
`turtle approximation', because with some imagination the right-most diagram
in fig.~\ref{fig:Gamma} resembles a turtle.
Here and in the following we again suppress the channel indices for brevity, and $p$ is 
the momentum of the incoming baryon.
Using the explicit form of $T$ given in eq.~(\ref{eq:Tdecomp}), we find
\begin{align}
\Gamma(\slashed{q},\slashed{p}) &= \slashed{q}\gamma_{5}\hat g 
 + T(\slashed{q},\slashed{p}-m;p)[(\slashed{p}-m)G(p)-I_{\mathrm{M}}] \gamma_{5}\hat g\no \\
&= [\slashed{q}\slashed{p}\Gamma_{1}(p)+\slashed{q}\Gamma_{2}(p)
 +  \slashed{p}\Gamma_{3}(p)+\Gamma_{4}(p)]\gamma_{5}
\end{align}
with scalar coefficient functions $\Gamma_{n}$ given by 
\begin{eqnarray}
\Gamma_{1}(p) &=& T_{1}(p^{2}H_{1}-mH_{0})+T_{2}(H_{0}-mH_{1})\no\\
 &&+T_{4}H_{0}+T_{6}H_{1} \,, \\
\Gamma_{2}(p) &=& \hat
g+p^{2}T_{1}(H_{0}-mH_{1})+T_{2}(p^{2}H_{1}-mH_{0})\no\\ 
&&+p^{2}T_{4}H_{1}+T_{6}H_{0} \,, \\
\Gamma_{3}(p) &=& T_{3}(p^{2}H_{1}-mH_{0})+T_{5}(H_{0}-mH_{1})\no\\ 
&&+T_{7}H_{0}+T_{8}H_{1} \,, \\
\Gamma_{4}(p) &=& p^{2}T_{3}(H_{0}-mH_{1})+T_{5}(p^{2}H_{1}-mH_{0})\no\\
&&+p^{2}T_{7}H_{1}+T_{8}H_{0} \,, 
\end{eqnarray}
where
\begin{eqnarray*}
H_{1} &=& (G_{0}(p)-mG_{1}(p))\hat g \,, \\
H_{0} &=& (p^{2}G_{1}(p)-mG_{0}(p)-I_{\mathrm{M}})\hat g \,.
\end{eqnarray*}
The functions $G_{0,1}(p)$ have been defined in eq.~(\ref{eq:G01}).

In order to complete step (1), we still have to specify which meson-baryon 
channels contribute in the framework of our electroproduction model. 
In this first study, we choose to consider only the ground-state octets of mesons 
and baryons, respectively. Moreover, from the topology of the hadronic part of 
the Feynman graph in fig.~\ref{fig:Gamma}
we can conclude that the meson-baryon pairs must have the charge and 
strangeness quantum numbers of the proton. This limits the number of channels 
to six: 
\beq
p\pi^{0},\ n\pi^{+},\ p\eta,\ \Lambda K^{+},\ \Sigma^{0}K^{+},\ \Sigma^{+}K^{0} \,.
\eeq
The limitation to these channels can only occur because our amplitude is not 
crossing-symmetric, otherwise more channels with different quantum numbers must 
be considered. The violation of crossing-symmetry is a drawback of the BSE method
which we use to iterate the rescattering graphs. To our knowledge, the
implementation of both crossing symmetry and exact (two-body) unitarity on the
basis of Feynman diagrams has not yet been accomplished. Other methods such
as, {\it e.g.}, an analog of the Roy equations (or generalizations thereof) for 
pion-pion, pion-kaon or pion-nucleon scattering, see 
\cite{Ananthanarayan:2000ht,Buettiker:2003pp,Becher:2001hv},
might be required to achieve this. Here, our goal is more modest,
and we sacrifice crossing symmetry in favor of unitarity. 

\begin{figure*}[t]
\centerline{\includegraphics[width=10.5cm]{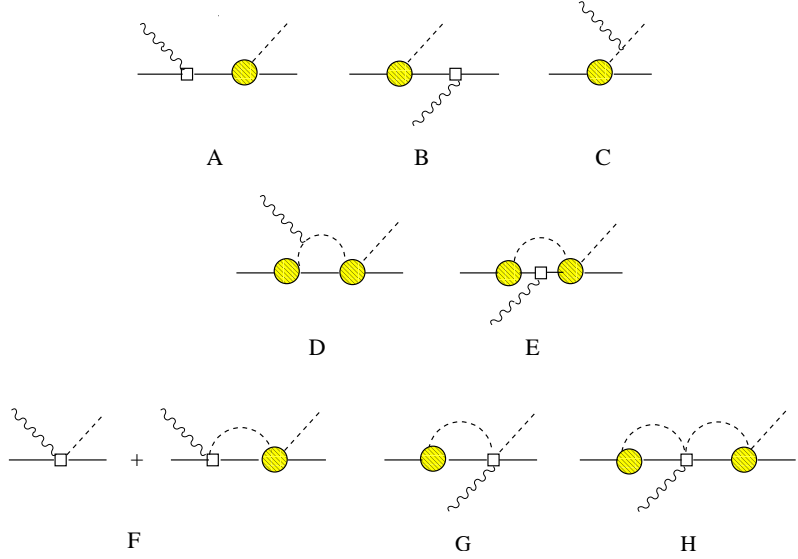}}
\caption{Classes of diagrams for kaon production off the nucleon in the 
turtle approximation utilized here.}
\label{Turtle}
\end{figure*}

By now, we have finished the first part (step~(1)) of our program. Our next task 
is to couple the photon to the hadronic part of the amplitude in a gauge-invariant 
fashion. Inserting the photon coupling at every possible place leads to the set of  
diagrams displayed in fig.~\ref{Turtle}.

The first three graphs are generated by coupling the photon to the external 
lines of the hadronic part. This leads to the expressions
\beqa
\label{eq:s} S_{s}^{\mu} &=& \Gamma(\slashed{q},\slashed{p})iS(\slashed{p})(ie\gamma^{\mu}) \,,\\   
\label{eq:u} S_{u}^{\mu} &=& (ieQ_{\mathrm{B}}\gamma^{\mu})iS(\slashed{p}_{1}-\slashed{q})
                             \Gamma(\slashed{q},\slashed{p}_{1}) \,,\\                              
\label{eq:t} S_{t}^{\mu} &=& ieQ_{\mathrm{M}}(2q-k)^{\mu}i\Delta(q-k)
                             \Gamma(\slashed{q}-\slashed{k},\slashed{p}_{1}) \,.                    
\eeqa
Here and in the following, $k$ is the four-momentum of the incoming photon, $q$ is the 
four-momentum of the outgoing Goldstone boson and $p_{1}$ and $p_{2}$ are the four-momenta 
of the incoming and outgoing baryon, respectively, while $p\equiv p_{1}+k$.
We have also introduced the meson (M) and baryon (B) charge matrices. They are 
diagonal matrices in channel space and read 
\begin{eqnarray*}
Q_{\mathrm{M}} &=& \diag(0,1,0,1,1,0) \,, \\
Q_{\mathrm{B}} &=& \diag(1,0,1,0,0,1) \,.
\end{eqnarray*}
The expressions for the three amplitudes $S_{s,u,t}^{\mu}$ must be decomposed into 
their independent Lorentz structures which is outlined in App.~\ref{app:decomp}.
Next we consider the coupling of the photon to internal lines of the bubble chain, 
leading to those diagrams that are most tedious to work out.
Consider first the coupling of the photon to an internal baryon line, see
fig.~\ref{Turtle}\,E. Using the pertinent Feynman rules, the 
contribution of this graph is easily seen to be
\begin{multline}
S_{\mathrm{B}}^{\mu} = -i\int\frac{d^{d}l}{(2\pi)^{d}}T(\slashed{q},\slashed{l};p)
              S(\slashed{p}-\slashed{l})\Delta(l)eQ_{\mathrm{B}}\gamma^{\mu} \\
\times S(\slashed{p}_{1}-\slashed{l})\Gamma(\slashed{l},\slashed{p}_{1}) \,,
\end{multline}
while the coupling of the photon to an internal meson line (fig.~\ref{Turtle}\,D) leads to
\begin{multline}
S_{\mathrm{M}}^{\mu} = -i\int\frac{d^{d}l}{(2\pi)^{d}}T(\slashed{q},\slashed{p}-\slashed{l};p)
                       S(\slashed{l})\Delta(p-l)eQ_{\mathrm{M}} \\
\times(2(p_{1}-l)+k)^{\mu} \Delta(p_{1}-l)\Gamma(\slashed{p}_{1}-\slashed{l},\slashed{p}_{1}) \,.
\end{multline}
Again, the decomposition into independent Lorentz structures can be found in
App.~\ref{app:decomp}. 

The next class of graphs we consider arises due to the `Kroll-Ruderman' (KR) 
term stemming from the covariant derivative $\nabla_{\mu}U$ in the chiral Lagrangian.
The corresponding tree level vertex reads (in matrix form)
\beq
S_{\textrm{KR,tree}}^{\mu} = eQ_{\mathrm{M}}\, \hat g \, \gamma^{\mu}\gamma_{5} \,.
\eeq
We add the loop contribution that accounts for the final-state interaction
after the meson has left the KR vertex, to arrive at (see fig.~\ref{Turtle}\,F)
\beq
S_{\textrm{KR}}^{\mu} = S_{\textrm{KR,tree}}^{\mu} +
\int\frac{d^{d}l}{(2\pi)^{d}}T(\slashed{q},\slashed{l};p)iS(\slashed{p}-\slashed{l})
\Delta(l)S_{\textrm{KR,tree}}^{\mu} \,.
\eeq

The remaining graphs are depicted in figs.~\ref{Turtle}\,G,H. 
These contributions arise from the terms proportional to the external vector
field in the chiral connection $\Gamma^{\mu}$, leading to a $\bar B\phi\phi\gamma B$ vertex rule
\beq
-ie\, \gamma^{\mu} \{Q_{\mathrm{M}}, g\}~,
\eeq
and to the following expressions for the Feynman graphs of fig.~\ref{Turtle}\,G,H:
\begin{align}
S_{\textrm{WT1}}^{\mu} &= e\gamma^{\mu}\{Q_{\mathrm{M}},g\} \int\frac{d^{d}l}{(2\pi)^{d}}
                          iS(\slashed{p}_{1}-\slashed{l}) \Delta(l) \Gamma(\slashed{l},\slashed{p}_{1}) \,, \\
S_{\textrm{WT2}}^{\mu} &= \int\frac{d^{d}\tilde l}{(2\pi)^{d}} T(\slashed{q},\slashed{\tilde{l}};p)
                          iS(\slashed{p}-\slashed{\tilde{l}})\Delta(\tilde l)S_{\textrm{WT1}}^{\mu} \,.
\end{align}
The Lorentz structure decomposition of all the above diagrams and the according 
contributions to the invariant amplitudes $B_{i}$ are given in App.~\ref{app:decomp}.

\begin{figure}[t]
\includegraphics[width=8.5cm]{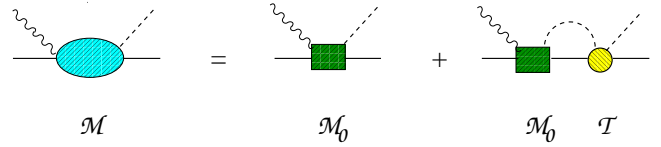}
\caption{Illustration of the integral equation (\ref{eq:inteq}) satisfied by our electroproduction amplitude}
\label{IntEq}
\end{figure}

Having finished the construction of the electroproduction amplitude, we return to the 
issue of unitarity. The crucial observation here is that every electroproduction amplitude 
$\mathcal{M}^{\mu}(q,k;p)$ which may be written as (see also fig.~\ref{IntEq})
\begin{align} \label{eq:inteq}
\mathcal{M}^{\mu}(q,k;p) &= \mathcal{M}^{\mu}_{0}(q,k;p) \no \\ 
&+ \int\frac{d^{d}l}{(2\pi)^{d}}\mathcal{T}(q,l;p)iS(\slashed{p}-\slashed{l})
   \Delta(l)\mathcal{M}^{\mu}_{0}(l,k;p)
\end{align}
obeys the requirement of two-body unitarity in the subspace of meson-baryon channels.
Here, $\mathcal{T}$ is an amplitude for meson-baryon scattering that solves a BSE of 
the type of eq.~(\ref{eq:BSE}) and $\mathcal{M}^{\mu}_{0}$ is a real kernel.

The proof proceeds in close analogy to the one presented in sect.~5 of \cite{Borasoy:2005zg}.
Now we note that our electroproduction amplitude given above decomposes into
five `unitarity classes', each of which obey an integral 
equation of the above type (\ref{eq:inteq}) by construction (with $\mathcal{T} = T$):
\begin{eqnarray*}
{\textrm{Class~1}} &:& S_{s}^{\mu} \\
{\textrm{Class~2}} &:& S_{u}^{\mu}+S_{\mathrm{B}}^{\mu} \\
{\textrm{Class~3}} &:& S_{t}^{\mu}+S_{\mathrm{M}}^{\mu} \\
{\textrm{Class~4}} &:& S_{\textrm{KR}}^{\mu} \\
{\textrm{Class~5}} &:& S_{\textrm{WT1}}^{\mu}+S_{\textrm{WT2}}^{\mu} .
\end{eqnarray*}
Hence, each class leads to a electroproduction amplitude that obeys unitarity for 
itself, though not gauge invariance.
This is because the expressions $S_{\textrm{WT1}}^{\mu}$, $S_{t}^{\mu}$ and $S_{u}^{\mu}$ 
as well as the tree graphs of $S_{\textrm{KR}}^{\mu}$ and $S_{s}^{\mu}$ are real in the 
physical region for the electroproduction process and can therefore constitute a kernel 
$\mathcal{M}^{\mu}_{0}$ in eq.~(\ref{eq:inteq}). The sum
\beq \label{eq:Msum}
\mathcal{M}^{\mu} \equiv S_{s}^{\mu}+S_{u}^{\mu}+S_{t}^{\mu}+S_{\mathrm{B}}^{\mu}+S_{\mathrm{M}}^{\mu}
                        +S_{\textrm{KR}}^{\mu}+S_{\textrm{WT1}}^{\mu}+S_{\textrm{WT2}}^{\mu}
\eeq
will then also obey unitarity, due to the linearity of the integral equation (\ref{eq:inteq}) 
in $\mathcal{M}^{\mu}_{0}$.

We will now show that the sum $\mathcal{M}^{\mu}$ of the five classes is gauge 
invariant by proving $k\cdot \mathcal{M} = 0$ for on-shell mesons and baryons. 
This might be obvious from our construction, but we include the proof for completeness.
The contraction of $k$ with the different amplitudes yields
\begin{align*}
k\cdot S_{s} &= -e\Gamma(\slashed{q},\slashed{p}) \,, \\
k\cdot S_{u} &= eQ_{\mathrm{B}}\Gamma(\slashed{q},\slashed{p}_{1}) \,, \\
k\cdot S_{t} &= eQ_{\mathrm{M}}\Gamma(\slashed{q}-\slashed{k},\slashed{p}_{1}) \,, \\ 
k\cdot S_{\mathrm{B}} &= \int\frac{d^{d}l}{(2\pi)^{d}}T(\slashed{q},\slashed{l};p)
                iS(\slashed{p}-\slashed{l})\Delta(l)eQ_{\mathrm{B}}
                \Gamma(\slashed{l},\slashed{p}_{1}) \\
             &- \int\frac{d^{d}l}{(2\pi)^{d}}T(\slashed{q},\slashed{l};p)
                eQ_{\mathrm{B}}iS(\slashed{p}_{1}-\slashed{l})\Delta(l)
                \Gamma(\slashed{l},\slashed{p}_{1}) \,, \\
k\cdot S_{\mathrm{M}} &= \int\frac{d^{d}l}{(2\pi)^{d}}T(\slashed{q},\slashed{l};p)
                iS(\slashed{p}-\slashed{l})\Delta(l)eQ_{\mathrm{M}} \\
             & \qquad \times \Gamma(\slashed{l}-\slashed{k},\slashed{p}_{1}) \\
             &- \int\frac{d^{d}l}{(2\pi)^{d}}T(\slashed{q},\slashed{l}+\slashed{k};p)
                eQ_{\mathrm{M}}iS(\slashed{p}_{1}-\slashed{l})\\
             & \qquad \times \Delta(l)\Gamma(\slashed{l},\slashed{p}_{1}) \,, \\
k\cdot S_{\textrm{KR}} &= eQ_{\mathrm{M}}\hat g\slashed{k}\gamma_{5} \\
             &+ \int\frac{d^{d}l}{(2\pi)^{d}}T(\slashed{q},\slashed{l};p)
                iS(\slashed{p}-\slashed{l})\Delta(l)eQ_{\mathrm{M}}\hat g\slashed{k}\gamma_{5} \,, \\
k\cdot S_{\textrm{WT1}} &= e\slashed{k}\{Q_{\mathrm{M}},g\}\int\frac{d^{d}l}{(2\pi)^{d}}
                iS(\slashed{p}_{1}-\slashed{l})\Delta(l)\Gamma(\slashed{l},\slashed{p}_{1}) \,, \\
k\cdot S_{\textrm{WT2}} &= \int\frac{d^{d}\tilde l}{(2\pi)^{d}}T(\slashed{q},\slashed{\tilde{l}};p)
                iS(\slashed{p}-\slashed{\tilde{l}})\Delta(\tilde l)e\slashed{k}\{Q_{\mathrm{M}},g\} \\
             & \qquad \times \int\frac{d^{d}l}{(2\pi)^{d}}iS(\slashed{p}_{1}-\slashed{l})\Delta(l)
                \Gamma(\slashed{l},\slashed{p}_{1}) \,.
\end{align*}
Here, charged external particles are put on-shell. First, we note that the tree graphs 
are gauge invariant for themselves, since the tree part of $k\cdot(S_{s}+S_{u}+S_{t})$ equals
\begin{displaymath}
-e\hat g\slashed{q}\gamma_{5} + eQ_{\mathrm{B}}\hat g\slashed{q}\gamma_{5}
+ eQ_{\mathrm{M}}\hat g(\slashed{q}-\slashed{k})\gamma_{5} \,,
\end{displaymath}
which is exactly canceled by the first term of $k\cdot S_{\textrm{KR}}$ (recall $Q_{\mathrm{B}}+Q_{\mathrm{M}}=1$).  
In order to deal with the loop contributions, it is useful to rewrite the integral 
equations (\ref{eq:BSE}) and (\ref{eq:Gamma}) for $T$ and $\Gamma$, respectively,
as
\begin{align} \label{eq:newBSE}
T(\slashed{q},\slashed{l};p) &= g(\slashed{q}+\slashed{l})\no\\
 &+ \int\frac{d^{d}\tilde l}{(2\pi)^{d}}T(\slashed{q},\slashed{\tilde{l}};p)
    iS(\slashed{p}-\slashed{\tilde{l}})\Delta(\tilde l)g(\slashed{\tilde l}+\slashed{l}) \,,
\end{align}
and
\beq \label{eq:newGamma}
\Gamma(\tilde{\slashed{l}},\slashed{p}_{1}) = \hat g\slashed{\tilde l}\gamma_5 
+ \int\frac{d^{d}l}{(2\pi)^{d}}g(\slashed{\tilde l}+\slashed{l})
  iS(\slashed{p}_{1}-\slashed{l})\Delta(l)\Gamma(\slashed{l},\slashed{p}_{1}) \,.
\eeq
These equations are equivalent to eqs.~(\ref{eq:BSE}) and (\ref{eq:Gamma}). 
Using eq.~(\ref{eq:newBSE}) in the second term of $k\cdot S_{\mathrm{B}}$ and $k\cdot S_{\mathrm{M}}$, 
and eq.~(\ref{eq:newGamma}) in the first term of $k\cdot S_{\mathrm{B}}$ and $k\cdot S_{\mathrm{M}}$, 
it is straightforward to show that the loop contributions in $k\cdot\mathcal{M}$ cancel. 
This, together with the above result for the tree contributions, completes the proof 
of gauge invariance for the full amplitude $\mathcal{M}$.

\section{Results}
\label{sec:results}
\def\theequation{\arabic{section}.\arabic{equation}}
\setcounter{equation}{0}

In this section, we will perform an overall $\chi^2$ fit to available 
photoproduction and pion-induced data on the proton near the respective thresholds. 
In more detail, we fit the differential cross sections for photoproduction 
on the proton into the
$K^+ \Lambda$, $K^+ \Sigma^0$, $K^0 \Sigma^+$ final states as well as of 
$\pi^- p \to K^0 \Lambda$, $K^0 \Sigma^0$.
Inspection of the differential cross sections reveals that already at moderate
energies away from threshold $p$- and $d$-waves become increasingly important.
Since our approach, which is based on the Weinberg-Tomozawa interaction kernel,
generates mainly  $s$-waves (with one important exception discussed below) 
and thus does not provide a realistic description
for higher partial waves, we expect it to be valid only in the near-threshold regions.
We have thus restricted our fits to energy values for which the differential
cross sections are dominated by $s$-waves,
 {\it i.e.}\ center-of-mass energies of about 1.80\,GeV corresponding to photon lab 
momenta of about 1.25\,GeV or pion lab momenta of about 1.23\,GeV. Still, we will be
able to extract interesting information from such investigations, in
particular, we can study in detail the commonly appearing approximations
made in the literature as already mentioned in the introduction.
The extension to higher energies requires inclusion of higher order counter terms
from the effective Lagrangian in the derivation of the interaction kernel.
However, this is beyond the scope of the present  work. In this investigation, 
we focus on the importance of gauge invariance and study the on-shell
approximation in the driving terms
and are satisfied with a moderate description of both photoproduction and pion-induced
data on the proton near threshold. Our investigation sets the stage for 
systematic improvements in the future which will lead to better agreement with
experiment.

\begin{figure*}[hthp]
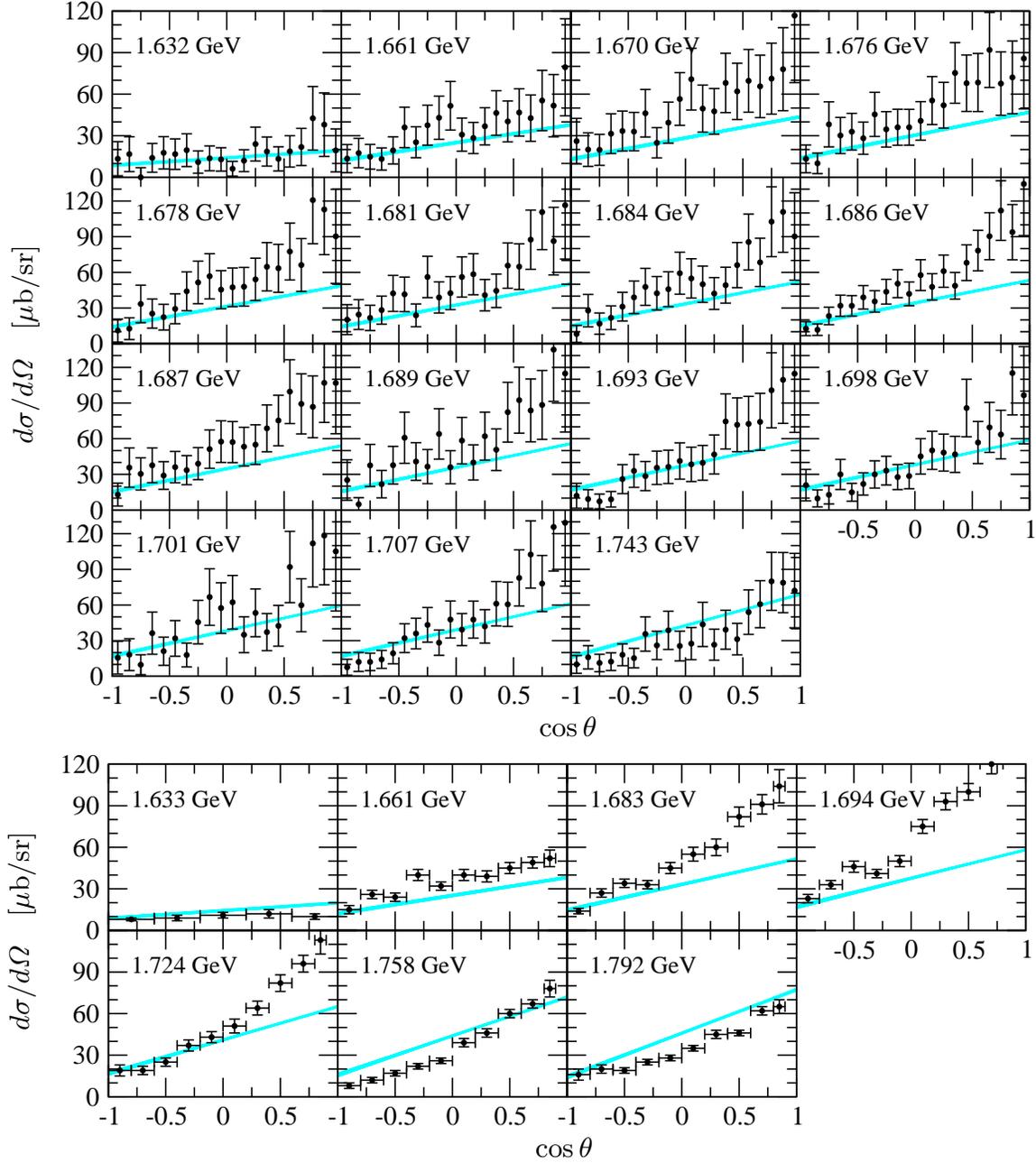

\centering
\begin{overpic}[width=0.8\textwidth,clip]{pldCSpipLamK.eps}
  \put(-5,27){\rotatebox{90}{\scalebox{1.3}{$d \sigma/d \Omega$\quad [$\mu$b/sr]}}}
  \put(49,-3){\scalebox{1.3}{$\cos \theta$}}
\end{overpic}\\[7mm]
\begin{overpic}[width=0.8\textwidth,clip]{pldCSpipLamB.eps}
  \put(-5,10){\rotatebox{90}{\scalebox{1.3}{$d \sigma/d \Omega$\quad [$\mu$b/sr]}}}
  \put(49,-3){\scalebox{1.3}{$\cos \theta$}}
\end{overpic}
\vspace*{7mm}
\caption{Differential cross sections for $\pi^- p \to K^0 \Lambda$ compared to data
         from \cite{Knasel:1975rr} (upper panel) and \cite{Baker:1978qm} (lower panel).
         The number in each plot denotes the respective c.m.\ energy $\sqrt{s}$.}
\label{fig:dCSpiLam}
\end{figure*}

The free parameters in our approach are, on the one hand, the 
three meson decay constants
$F_\pi$, $F_K$, $F_\eta$ which we vary separately within realistic bounds, 
as the SU(3) symmetry-breaking differences between them are 
beyond our working precision of the effective potential.
More precisely, SU(3) symmetry breaking is generated by various higher
order terms in the meson-baryon (meson)  Lagrang\-ian starting at 
chiral order two (four). Since these contributions 
are not included in the leading order WT
kernel, we simulate such effects by allowing variations in the various
meson decays constants. This, of course, will no longer be done when
the higher order terms in the interaction kernel have been included. 
One observes that the fitted decay constants tend to larger values
reducing the strength of the Weinberg-Tomozawa interaction. This
is consistent with findings in earlier chiral unitary studies, {\it i.e.}\  the
WT interaction  in many cases produces too strong $s$-waves, see {\it e.g.} \cite{Borasoy:2002mt}.
We allow $F_\pi, F_K,$ and $F_\eta$ to vary between 70\,MeV and 150\,MeV.
The values we find for the best fits, {\it i.e.}\ the fits with the lowest 
overall $\chi^2$ values, are
\beqa 
F_\pi   &=& (113 \ldots 127)\,\textrm{MeV}~, \no\\ 
F_K     &=& (149 \ldots 150)\,\textrm{MeV}~, \no\\
F_\eta  &=& ( 74 \ldots  82)\,\textrm{MeV}~. 
\eeqa
Note that $F_K$ tends towards the upper limit of 150\,MeV.
On the other hand, we fit the four different isospin-sym\-met\-ric 
scales $\mu$ in the loop integrals. For the best fits their values are:
\beqa
\mu_{\pi N}     &=& (0.46 \ldots 0.54)\,\textrm{GeV}~,\no\\
\mu_{\eta N}    &=& (3.29 \ldots 4.41)\,\textrm{GeV}~,\no\\
\mu_{K \Lambda} &=& (2.56 \ldots 2.86)\,\textrm{GeV}~,\no\\
\mu_{K \Sigma}  &=& (3.66 \ldots 4.31)\,\textrm{GeV}~.
\eeqa
We note that these values are roughly in accordance
with the {\it natural size} estimate of ref.~\cite{Meissner:1999vr}
(although most of them turn out to be somewhat large).

\begin{figure}
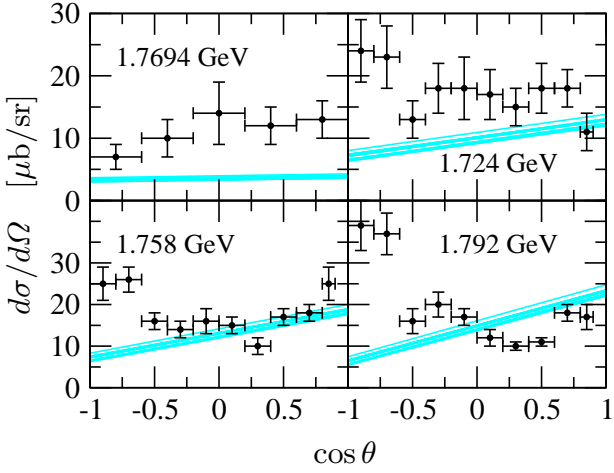

\centering
\begin{overpic}[width=0.42\textwidth,clip]{pldCSpipSig0B.eps}
  \put(-7,19){\rotatebox{90}{\scalebox{1.3}{$d \sigma/d \Omega$\quad [$\mu$b/sr]}}}
  \put(47,-7){\scalebox{1.3}{$\cos \theta$}}
\end{overpic}
\vspace*{6mm}
\caption{Differential cross sections for $\pi^- p \to K^0 \Sigma^0$ compared to data
         from \cite{Baker:1978bb}.
         The number in each plot denotes the respective c.m.\ energy $\sqrt{s}$.}
\label{fig:dCSpiSig0}
\end{figure}

\begin{figure}
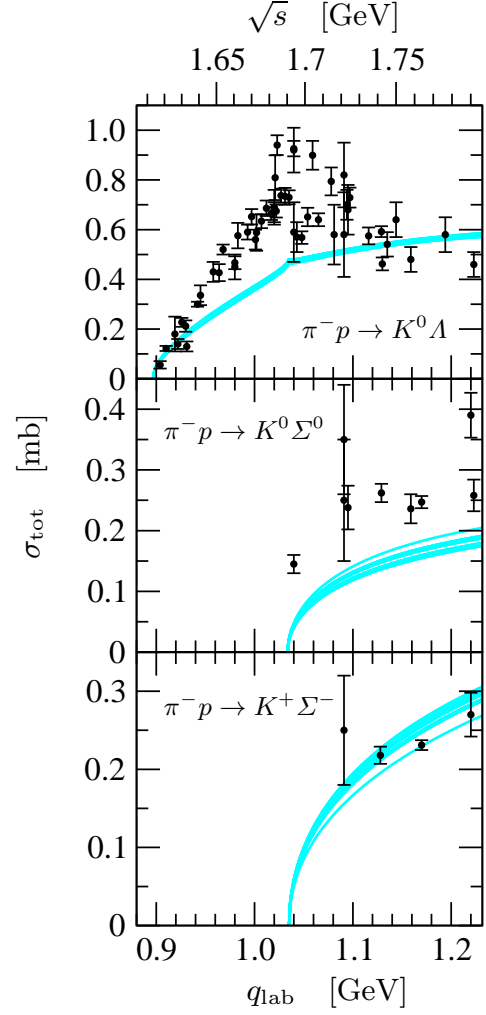

\centering
\vspace*{6mm}
\begin{overpic}[width=0.30\textwidth,clip]{pltCSpi.eps}
  \put(-7,44){\rotatebox{90}{\scalebox{1.3}{$\sigma_{\text{tot}}$\quad [mb]}}}
  \put(18,-4){\scalebox{1.3}{$q_{\text{lab}}$\quad [GeV]}}
  \put(18,102){\scalebox{1.3}{$\sqrt{s}$\quad [GeV]}}
  \put(24,68){\scalebox{1.1}{$\pi^- p \to K^0 \Lambda$}}
  \put(9,57){\scalebox{1.1}{$\pi^- p \to K^0 \Sigma^0$}}
  \put(9,27){\scalebox{1.1}{$\pi^- p \to K^+ \Sigma^-$}}
\end{overpic}
\vspace*{7mm}
\caption{Total cross sections for $\pi^-$ induced production of $K^0 \Lambda$ (top), 
         $K^0 \Sigma^0$ (middle), $K^+ \Sigma^-$ (bottom) as a function of the pion
         lab momentum $q_{\text{lab}}$. The corresponding c.m.\ energy
         $\sqrt{s}$ is also given.  The data are taken from 
         \cite{Knasel:1975rr,Baker:1978qm,Baker:1978bb} and the compilation
         \cite{Landolt_Boernstein}.}
\label{fig:tCSpi}
\end{figure}

In figs.~\ref{fig:dCSpiLam}, \ref{fig:dCSpiSig0} we present the best fit results 
for the pion-induced differential cross sections 
$\pi^- p \to K^0 \Lambda$, $K^0 \Sigma^0$, respectively.
Although the WT interaction kernel is entirely $s$-wave, there are
$p$-wave contributions due to the lower components of the Dirac spinors 
which are introduced in the calculation of the scattering matrix $T$.
The corresponding total cross sections and the one for
$\pi^- p \to K^+ \Sigma^-$ are shown in fig.~\ref{fig:tCSpi}. We remark
that the bands in these figures are generated by about 20 fits with
a $\chi^2$ very close to its minimal value. Since with the simple
WT interaction we are not able to describe all these and the
photon-induced data to a high accuracy, we refrain from giving 
one-sigma error bands as done in our study on $K^- p$ scattering
\cite{Borasoy:2006sr}. This will be done in future work when the
higher order terms in the interaction kernel will be included.

The processes $\pi^- p \to K^+ \Sigma^-$ and $\pi^+ p \to K^+ \Sigma^+$ are
dominated by $p$- and $d$-waves already close to threshold.
Hence, a realistic description of these channels cannot be provided by the leading 
WT interaction. 
In order not to overestimate the data we have still included the total 
cross section for $\pi^- p \to K^+ \Sigma^-$ in our fit.
We refrain, however, from taking into account data on $\pi^+ p \to K^+ \Sigma^+$
at all since, 
regardless of the choice of parameters (within realistic ranges),
our approach clearly overshoots the  $\pi^+ p \to K^+ \Sigma^+$ cross section.
This indicates that higher order contact interactions in the interaction kernel are
absolutely necessary---particularly for this process---in order to reduce the strength of the WT term.
Moreover, it is worthwhile mentioning that the presented fits exhibit a peak in the
$\pi^- p \to \eta n$ cross section due to the $S_{11}(1535)$ resonance. 
Pion- and photon-induced eta production data will 
be included in future work after the interaction kernel has been developed to higher chiral orders.

\begin{figure*}[p]
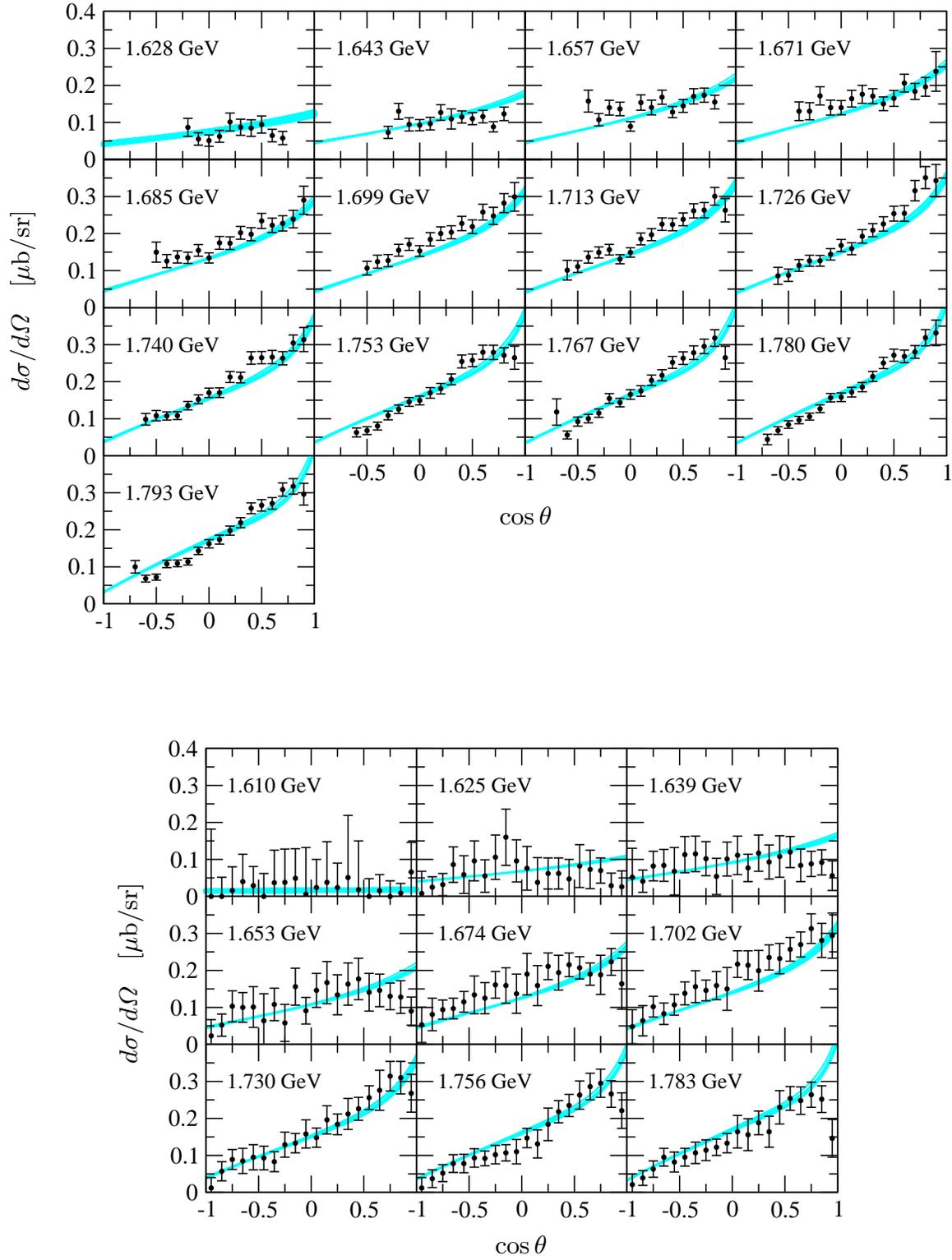

\centering
\begin{overpic}[width=0.8\textwidth,clip]{pldCSphLamC.eps}
  \put(-6,27){\rotatebox{90}{\scalebox{1.3}{$d \sigma/d \Omega$\quad [$\mu$b/sr]}}}
  \put(49,12){\scalebox{1.3}{$\cos \theta$}}
\end{overpic}\\[17mm]
\begin{overpic}[width=0.61\textwidth,clip]{pldCSphLamS.eps}
  \put(-7,23){\rotatebox{90}{\scalebox{1.3}{$d \sigma/d \Omega$\quad [$\mu$b/sr]}}}
  \put(48.5,-5){\scalebox{1.3}{$\cos \theta$}}
\end{overpic}
\vspace*{15mm}
\caption{Differential cross sections for $\gamma p \to K^+ \Lambda$ compared to data
         from \cite{Bradford:2005pt} (upper panel) and \cite{Glander:2003jw} (lower panel).
         The number in each plot denotes the respective c.m.\ energy $\sqrt{s}$.}
\label{fig:dCSphLam}
\end{figure*}

\begin{figure}[h]
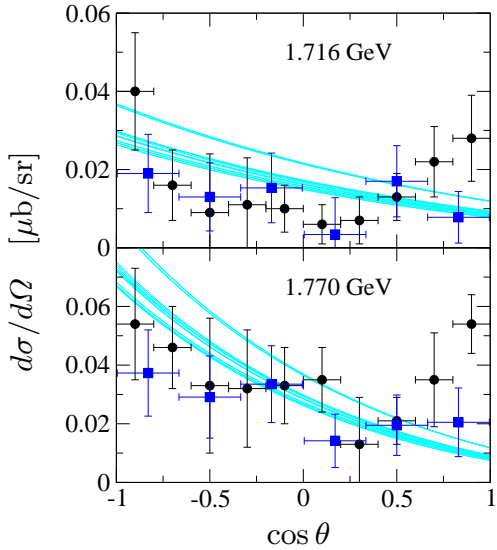

\centering
\begin{overpic}[width=0.32\textwidth,clip]{pldCSphSigp.eps}
  \put(-11,27){\rotatebox{90}{\scalebox{1.3}{$d \sigma/d \Omega$\quad [$\mu$b/sr]}}}
  \put(40,-7){\scalebox{1.3}{$\cos \theta$}}
\end{overpic}
\vspace*{6mm}
\caption{Differential cross sections for $\gamma p \to K^0 \Sigma^+$ compared to data
         from \cite{Lawall:2005np} (circles) and \cite{Castelijns:2007qt} (squares).
         The number in each plot denotes the respective c.m.\ energy $\sqrt{s}$.}
\label{fig:dCSphSigp}
\end{figure}

\begin{figure}[h]
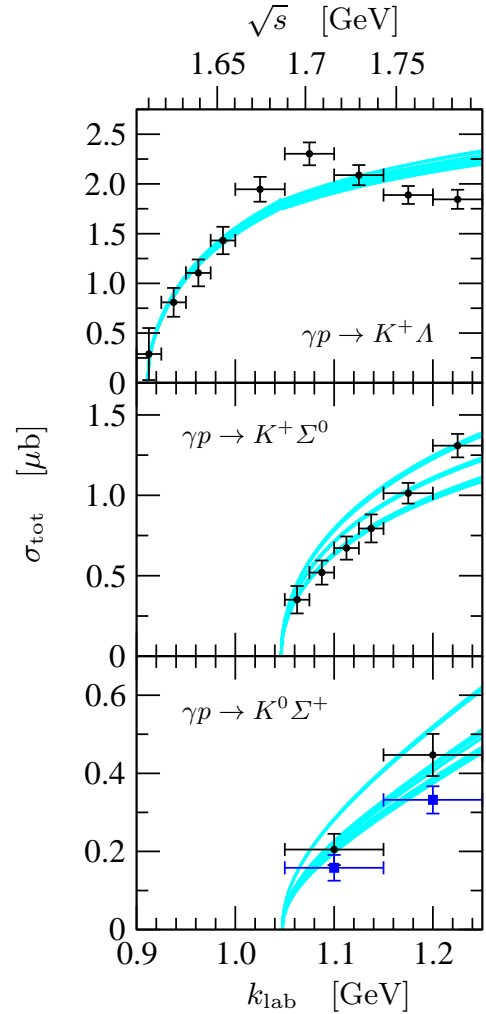

\centering
\vspace*{6mm}
\begin{overpic}[width=0.30\textwidth,clip]{pltCSph.eps}
  \put(-7,44){\rotatebox{90}{\scalebox{1.3}{$\sigma_{\text{tot}}$\quad [$\mu$b]}}}
  \put(18,-4){\scalebox{1.3}{$k_{\text{lab}}$\quad [GeV]}}
  \put(18,102){\scalebox{1.3}{$\sqrt{s}$\quad [GeV]}}
  \put(24,68){\scalebox{1.1}{$\gamma p \to K^+ \Lambda$}}
  \put(11,57){\scalebox{1.1}{$\gamma p \to K^+ \Sigma^0$}}
  \put(11,27){\scalebox{1.1}{$\gamma p \to K^0 \Sigma^+$}}
\end{overpic}
\vspace*{7mm}
\caption{Total cross sections for photoproduction of $K^+ \Lambda$ (top), 
         $K^+ \Sigma^0$ (middle), $K^+ \Sigma^0$ (bottom) as a function of the photon
         lab momentum $k_{\text{lab}}$. The corresponding c.m.\ energy
         $\sqrt{s}$ is also given.  The data are taken from \cite{Lawall:2005np} (circles)
         and \cite{Castelijns:2007qt} (squares).}
\label{fig:tCSph}
\end{figure}

Furthermore, we have included in our fit data on 
differential cross sections of the photoproduction processes
$\gamma p \to K^+ \Lambda$, $K^+ \Sigma^0$, $K^0 \Sigma^+$.
The results are displayed in figs.~\ref{fig:dCSphLam}, \ref{fig:dCSphSig0}, \ref{fig:dCSphSigp},
respectively, while the corresponding
total cross sections are presented in fig.~\ref{fig:tCSph}.
A few remarks are in order:  The SAPHIR and CLAS data on charged kaon
photoproduction  show some inconsistencies at forward angles, but this can
not be resolved within the approximations made here. 
Also, the very different shape of the differential cross sections
for the $K^+ \Lambda$ and $K^+ \Sigma^0$ final states can be traced
back to the isospin selectivity of the $\Lambda$, 
see also ref.~\cite{Anisovich:2007bq}.

\begin{figure*}
\centering
\begin{overpic}[width=0.8\textwidth,clip]{pldCSphSig0C.eps}
  \put(-6,9){\rotatebox{90}{\scalebox{1.3}{$d \sigma/d \Omega$\quad [$\mu$b/sr]}}}
  \put(49,-3.5){\scalebox{1.3}{$\cos \theta$}}
\end{overpic}\\[8mm]
\begin{overpic}[width=0.61\textwidth,clip]{pldCSphSig0S.eps}
  \put(-7,12){\rotatebox{90}{\scalebox{1.3}{$d \sigma/d \Omega$\quad [$\mu$b/sr]}}}
  \put(49,-4){\scalebox{1.3}{$\cos \theta$}}
\end{overpic}
\vspace*{6mm}
\caption{Differential cross sections for $\gamma p \to K^+ \Sigma^0$ compared to data
         from \cite{Bradford:2005pt} (upper panel) and \cite{Glander:2003jw} (lower panel).
         The number in each plot denotes the respective c.m.\ energy $\sqrt{s}$.}
\label{fig:dCSphSig0}
\end{figure*}

To summarize, the results presented here show a reasonable agreement with data
on  photon- and pion-induced reactions close to threshold, but more realistic 
interaction kernels and higher partial waves
are required to obtain better agreement with data, in particular away from threshold.
This is however beyond the scope of the present investigation.

\begin{figure*}
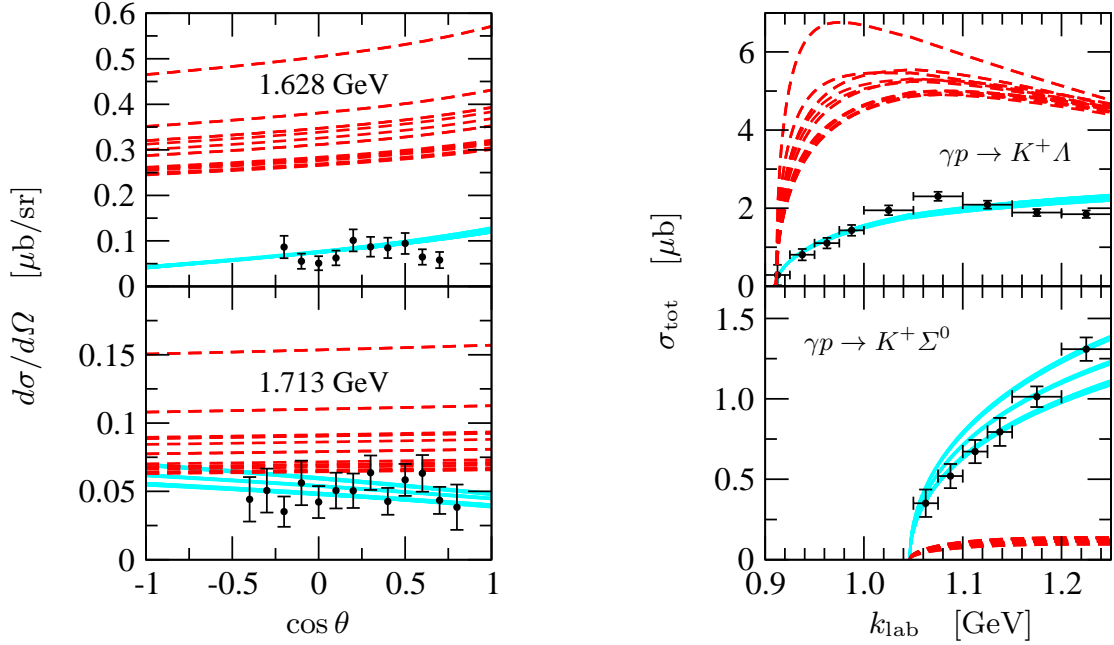

\centering
\begin{tabular}{cp{20mm}c}
\begin{overpic}[width=0.313\textwidth,clip]{pldCSph_AF.eps}
  \put(-11,32){\rotatebox{90}{\scalebox{1.3}{$d \sigma/d \Omega$\quad [$\mu$b/sr]}}}
  \put(35,-6){\scalebox{1.3}{$\cos \theta$}}
\end{overpic} & &
\begin{overpic}[width=0.30\textwidth,clip]{pltCSph_AF.eps}
  \put(-10,42){\rotatebox{90}{\scalebox{1.3}{$\sigma_{\text{tot}}$\quad [$\mu$b]}}}
  \put(27,-6){\scalebox{1.3}{$k_{\text{lab}}$\quad [GeV]}}
  \put(39,74){\scalebox{1.1}{$\gamma p \to K^+ \Lambda$}}
  \put(16,42){\scalebox{1.1}{$\gamma p \to K^+ \Sigma^0$}}
\end{overpic}
\end{tabular}
\vspace*{5mm}
\caption{Sample results which illustrate the importance of contributions which are needed to
         fulfill gauge invariance.
         Left panel: differential cross sections for photoproduction of $K^+ \Lambda$ (top), 
         $K^+ \Sigma^0$ (bottom) at c.m.\ energies stated in the figures compared to
         data from \cite{Bradford:2005pt}. 
         Right panel: total cross sections for photoproduction of $K^+ \Lambda$ (top),
         $K^+ \Sigma^0$ (bottom) compared to data from \cite{Glander:2003jw}.
         The full result is represented by the solid lines, whereas the dashed lines
         indicate the contributions from diagrams A+F in fig.~\ref{Turtle}.}
\label{fig:CSph_AF}
\end{figure*}

Of interest here are
gauge invariance violations encountered in previous chiral unitary approaches
which only took a subset of the diagrams in fig.~\ref{Turtle} into account. 
In order to estimate the typical size of gauge symmetry violations,
we omit all contributions where the photon couples to intermediate and final states
and retain only those diagrams where the photon is initially absorbed (figs.~\ref{Turtle}\,A+F).
Note that both the graphs A and F as well as their sum are unitary
as explained in detail in sect.~\ref{sec:photo}. For making this comparison, we
do not refit the parameters but use the values obtained in the full approach.
Some sample results are compared to the full approach in fig.~\ref{fig:CSph_AF}.
One clearly observes that violations of gauge invariance are sizable,
although this effect could, in principle, be concealed numerically by readjusting 
the parameters of the approach.
This indicates that the additional contributions which render the
photoproduction amplitudes gauge invariant and were omitted in previous work 
are not negligible and must be taken into account.

\begin{figure*}
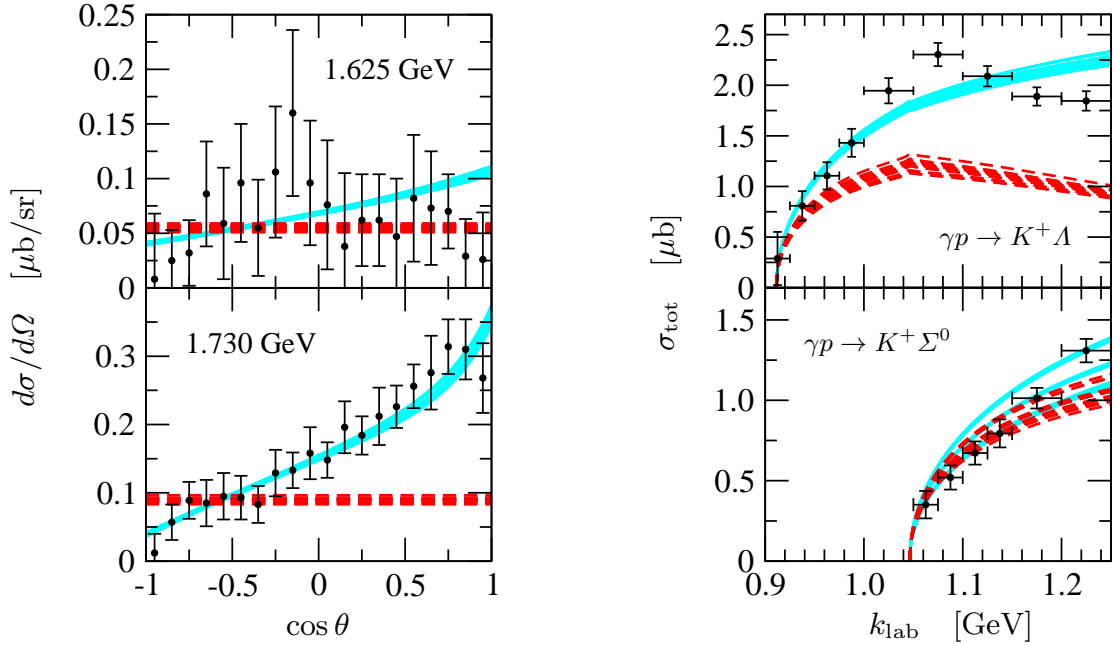

\vspace{1.3cm}
\centering
\begin{tabular}{cp{20mm}c}
\begin{overpic}[width=0.313\textwidth,clip]{pldCSph_on.eps}
  \put(-11,32){\rotatebox{90}{\scalebox{1.3}{$d \sigma/d \Omega$\quad [$\mu$b/sr]}}}
  \put(35,-6){\scalebox{1.3}{$\cos \theta$}}
\end{overpic} & &
\begin{overpic}[width=0.30\textwidth,clip]{pltCSph_on.eps}
  \put(-10,42){\rotatebox{90}{\scalebox{1.3}{$\sigma_{\text{tot}}$\quad [$\mu$b]}}}
  \put(27,-6){\scalebox{1.3}{$k_{\text{lab}}$\quad [GeV]}}
  \put(39,60){\scalebox{1.1}{$\gamma p \to K^+ \Lambda$}}
  \put(16,42){\scalebox{1.1}{$\gamma p \to K^+ \Sigma^0$}}
\end{overpic}
\end{tabular}
\vspace*{5mm}
\caption{Sample results which illustrate the importance of contributions
         neglected in conventional coupled-channels analyses. 
         Left panel: differential cross sections for photoproduction of $K^+ \Lambda$
         at two different c.m.\ energies as stated in the figures. 
         Right panel: total cross sections for photoproduction of $K^+ \Lambda$ (top),
         $K^+ \Sigma^0$ (bottom).
         The full result is represented by the solid lines, whereas the dashed lines
         correspond to the outcome of the model \cite{Borasoy:2002mt} given the same 
         values of the fit parameters. 
         The data are taken from \cite{Glander:2003jw}.}
\label{fig:CSph_on}
\end{figure*}

In order to be able to compare our results with previous chiral unitary approaches
we have also worked in the approximations employed in these investigations,
see {\it e.g.}\  \cite{Borasoy:2002mt}\footnote{Note, however, that in ref.~\cite{Borasoy:2002mt}
the primary goal was the consistent inclusion of the $\eta'$ meson in chiral unitary
approaches.}.
To this aim, the interaction kernel for meson-baryon scattering 
is directly sandwiched between Dirac spinors and projected
onto the $s$-wave which is then iterated to infinite order in a geometric
series. Furthermore, for the photoproduction process only the leading
$s$-wave (the so-called $E_{0+}$ multipole) is considered.
Obviously, this scheme produces pure $s$-waves and cannot reproduce 
the structures from higher partial waves in the differential cross sections, 
see fig.~\ref{fig:CSph_on}. One notes that these approximations can indeed 
be sizable. Of course, in ref.~\cite{Borasoy:2002mt} higher order terms were included
in the  $E_{0+}$ multipole, but that does not change the observations
just made. Again, by a suitable parameter refitting one might be able to describe
the total cross sections, but given the more sophisticated scheme developed
here, such approximations are no longer necessary.

Overall, we have illustrated that both the on-shell approximation and 
the omission of certain classes of diagrams which are required to fulfill 
gauge invariance constitute crude approximations utilized within chiral 
unitary approaches in the past.
Although these effects can numerically be concealed to a large extent
by readjusting the parameters, important theoretical constraints are not
met in this manner. On the other hand, the approach we have developed here is in 
accordance with these criteria. We stress again that our results also 
show that it is mandatory to go beyond the leading WT approximation in the
interaction kernel and to properly include higher partial waves in the
BSE.
Moreover, contributions from three-particle intermediate states, such as $\pi \pi N$,
might play a role in kaon photoproduction,
see {\it e.g.}\ \cite{Inoue:2001ip}, but are beyond the scope of the 
present investigation and have thus been discarded. For a method to 
incorporate such three-particle cuts, see ref.~\cite{Motzke:2002fn}.

\section{Conclusions and outlook}
\def\theequation{\arabic{section}.\arabic{equation}}
\setcounter{equation}{0}
\label{sec:conclusions}

The search for missing resonances is of great importance and actively being pursued at
various experimental facilities. A promising tool to discover new resonances
is the photoproduction of kaons on protons. A strong coupling 
to channels with open strangeness could reveal new resonances yet undetected
in previous experiments based on pion-nucleon physics.
Experiments with open strangeness are currently performed at ELSA, JLab and at SPring-8
providing a host of experimental data.
The obtained data must be met by theoretical approaches and yield a set of tight
constraints.   
In this work, a chiral unitary approach based on the combination of the chiral 
effective Lagrangian with a coupled-channels Bethe-Salpeter equation is presented.
The method is exactly unitary and satisfies gauge invariance. It improves previous
approaches in this field which employed only a subset of the Feynman diagrams
considered here.

We have fitted both the differential cross sections for photoproduction into
$K^+ \Lambda$, $K^+ \Sigma^0$, $K^0 \Sigma^+$ as well as of the 
meson-baryon scattering processes 
$\pi^- p \to K^0 \Lambda$, $K^0 \Sigma^0$, $K^+ \Sigma^-$. In the fits, we have restricted
ourselves to the threshold regions of the respective channels, as we cannot expect
to obtain a realistic description of these processes away from threshold
due to the increasing importance of higher partial waves absent in the Weinberg-Tomozawa
interaction kernel. In order to improve the agreement with experimental data,
the inclusion of higher chiral orders in the interaction kernel is necessary.
In charged kaon photoproduction, important contributions to the $p$ and higher
partial waves are generated by the kaon pole term, which is already included 
at the order we are working.

Our aim was to estimate, on the one hand, the size of gauge invariance 
violations introduced in previous cou\-pled-channels analyses by neglecting 
Feynman diagrams where the photon couples to intermediate states of the Bethe-Salpeter bubble chain.
On the other hand, the simplification of setting the interaction kernel
on-shell and performing $s$-wave projections on the meson-baryon and the 
photon-baryon subsystems is studied. 
Our investigation suggests that both approximations are not justified in the treatment
of photoproduction processes and lead to sizable changes in the results.
It is thus important to satisfy gauge invariance and include off-shell terms
in the effective potential.
The approach presented here fulfills these requirements in a minimal way, {\it i.e.}\ 
any subset of the included diagrams cannot be omitted as this would violate
either exact unitarity or gauge invariance.

There are three directions in which the approach presented here needs to be
improved: First, the inclusion of higher order terms in the interaction
kernel and the resulting better inclusion of $p$-waves and higher multipoles
is straightforward but tedious. This will allow to include also the
polarization data into the analysis.
In this context, we should also mention that
in the chiral unitary approach not all resonances are generated dynamically,
so in fact one might have to include explicit resonance fields in certain
channels to achieve a precise description. Such a method has already been
developed for elastic pion-nucleon scattering, see ref.~\cite{Meissner:1999vr}.
Second, one should also include pion and photon induced $\eta, \eta '$
production data as already accomplished in \cite{Borasoy:2002mt} 
and further constrain the scattering and production amplitudes
through matching to the two-flavor sector (as done {\it e.g.}\ in 
refs.~\cite{Frink:2004ic,Oller:2006jw}).
Last but not least, the violation of crossing symmetry 
due to the turtle approximation needs to be repaired. This could,  in principle, 
be done by formulating Roy-type equations, but is technically  involved.

\section*{Acknowledgements}

We thank Ulrike Thoma for useful discussions and for providing us with the
CB-ELSA/TAPS data. This work is supported in part by DFG 
(SFB/TR 16,``Subnuclear Structure of Matter'', and BO 1481/6-1)  and is part of the 
EU Integrated Infrastructure Initiative Hadron
Physics Project under contract number RII3-CT-2004-506078.

\begin{appendix}

\section{Bethe-Salpeter approach with off-shell kernel}
\label{app:offshell}
\def\theequation{\Alph{section}.\arabic{equation}}
\setcounter{equation}{0}

In this appendix, we present an alternative solution of the Bethe-Salpeter
equation with off-shell pieces in the interaction kernel \cite{Beisert:thesis,Djukanovic:2006xc}.
In general, the amplitude for a meson-baryon scattering process 
$B_a(p_a) + \phi_i(q_i) \to B_b(p_b) + \phi_j(q_j)$ 
can be written as a function $V(t-u, \slashed{p}, \slashed{p}_c - m_c, q_k^2 - M_k^2)$ 
with masses $m_c$ ($c = a, b$) and $M_k$ ($k = i, j$) for the baryons and mesons, respectively, 
the total momentum $p = p_a + q_i = p_b + q_j$
and the Mandelstam variables $t = (p_a - p_b)^2$, $u = (p_a - q_j)^2$.
It is understood that the off-shell terms $\slashed{p}_a -m_a$ have been moved
to the right, whereas the $\slashed{p}_b -m_b$ have been moved to the left. 
We assume that $V$ is analytic in the variable $t-u$. This is certainly
the case for all polynomial interactions which are derived from contact interactions.
The Taylor expansion of $V$ in $t-u$ reads 
\begin{align}
V(t-u) &= \sum_{i=0}^{\infty} (t-u)^i V_i  \no \\
       &= V_0 + (t-u) V_1 + (t-u)^2 V_2 + \ldots \ ,
\end{align}
where we have suppressed the dependence of $V$ and $V_i$ on $\slashed{p}$ and the 
off-shell pieces $\slashed{p}_c -m_c$ and $p_k^2 -M_k^2$ for brevity.
In the center-of-mass frame the variable $t-u$ is related to the scattering angle 
$\theta$ via
\beq
t-u = 4 |\mbox{\boldmath$p$}_a| |\mbox{\boldmath$p$}_b| \cos \theta + \frac{1}{p^2} \
       (p_a^2 - q_i^2) \ (p_b^2 - q_j^2)  \,,
\eeq
where $|\mbox{\boldmath$p$}_{a,b}|$ are the moduli of the c.m.\ three-momenta.
We define a set of orthogonal functions $J_l$ with $l = 0, 1, 2,\ldots$
(or $s, p, d, \ldots$) which are proportional to the Legendre polynomials 
in the center-of-mass scattering angle with the first few $J_l$ given by
\begin{align}
J_s &= 1 \,, \no \\
J_p &= |\mbox{\boldmath$p$}_a| |\mbox{\boldmath$p$}_b| \cos \theta =
    \frac{1}{4} (t-u) - \frac{1}{4p^2} \  (p_a^2 - q_i^2) \ (p_b^2 - q_j^2) \,, \no \\
J_d &= |\mbox{\boldmath$p$}_a|^2 |\mbox{\boldmath$p$}_b|^2 (\cos^2 \theta - \frac{1}{3})
     = J_p^2 - \frac{1}{3} \ |\mbox{\boldmath$p$}_a|^2 |\mbox{\boldmath$p$}_b|^2 \,, \no \\
J_f &= |\mbox{\boldmath$p$}_a|^3 |\mbox{\boldmath$p$}_b|^3 (\cos^3 \theta - \frac{3}{5} \cos \theta )
     = J_p^3 - \frac{3}{5} \ |\mbox{\boldmath$p$}_a|^2 |\mbox{\boldmath$p$}_b|^2  J_p \,.
\end{align}
The moduli of the three-momenta in the center-of-mass frame can be expressed in 
terms of Lorentz scalars as
\beqa
|\mbox{\boldmath$p$}_a|^2 &=& \frac{1}{4p^2} \  (p^2+ q_i^2 - p_a^2)^2 - q_i^2 \,, \no \\
|\mbox{\boldmath$p$}_b|^2 &=& \frac{1}{4p^2} \  (p^2+ q_j^2 - p_b^2)^2 - q_j^2 \,.
\eeqa
The expansion of the amplitude $V$ in $t-u$ can now be reformulated as an expansion in $J_l$,
$V= \sum_l V_l J_l$.
In order to keep the presentation simple, we restrict ourselves 
to the Weinberg-Tomozawa interaction from now on. The generalization to more
complicated vertex structures is straightforward.
For the Weinberg-Tomozawa term $V$ does not depend on $t-u$,
\beq
V = V_0 = V_s  \ ,
\eeq
and the division of $V$ into on- and off-shell pieces can be conveniently accomplished
by employing the matrix notation (for the reaction $ai \to bj$)
\beqa   \label{eq:partwave}
V_{s}^{bj,ai} &=& \mbox{\boldmath$u$}_{bj}^T \,
\breve{V}_{s}^{bj,ai}\, \mbox{\boldmath$u$}_{ai} \,, \\
\mbox{\boldmath$u$}_{ai}^{T}(\slashed{p}_a) &=& (1, \slashed{p}_a - m_a) \,, \no \\
\mbox{\boldmath$u$}_{bj}^{T}(\slashed{p}_b) &=& (1, \slashed{p}_b - m_b) \,, \no \\
\breve{V}_{s}^{bj,ai}(\slashed{p})          &=& g^{bj,ai} 
   \left(\begin{array}{cc} 2 \slashed{p} - m_a - m_b & \ -1 \\
                                 -1                  & \  0 \end{array}\right) \,,
\eeqa
where $g^{bj,ai}$ is the Weinberg-Tomozawa coupling for the channels under consideration
as defined in eq.~(\ref{eq:WTcoupling}).
The amplitude $V$ is utilized as the interaction kernel in the Bethe-Salpeter equation
for the scattering matrix $T$, {\it cf.}\ eq.~(\ref{eq:BSE}), 
\begin{multline}
T(\slashed{q}_{j},\slashed{q}_{i};p) = V(\slashed{q}_{j},\slashed{q}_{i}) \\
+ \int\frac{d^{d}l}{(2\pi)^{d}}V(\slashed{q}_{j},\slashed{l})
          iS(\slashed{p}-\slashed{l})\Delta(l)T(\slashed{l},\slashed{q}_{i};p) \,.
\end{multline}
For the decomposition into on- and off-shell pieces we make the same 
ansatz for the partial wave decomposition of the BSE solution $T$ 
as for $V$ in eq.~(\ref{eq:partwave}),
\beq
T_{s}^{bj,ai} = \mbox{\boldmath$u$}_{bj}^T \,\breve{T}_{s}^{bj,ai}\, \mbox{\boldmath$u$}_{ai} \,.
\eeq
This yields a BSE for $\breve{T}_{s}$ which is a matrix equation in the combined channel and 
off-shell space
\beq
\breve{T}_{s} = \breve{V}_{s} + \breve{V}_{s} \breve{G}_{s} \breve{T}_{s}
\eeq
with a matrix $\breve{G}_{s}$ which is diagonal in channel space with off-shell submatrices
\beq
\breve{G}_{s}^{ck,ck} = \int\frac{d^{d}l}{(2\pi)^{d}} \mbox{\boldmath$u$}_{ck}(\slashed{l})
                        iS(\slashed{l})\Delta(p-l) \mbox{\boldmath$u$}_{ck}^{T}(\slashed{l}) \,.
\eeq
Hence, the integral expression in the BSE factorizes also without putting the interaction kernel $V$ 
and the solution $T$ on-shell. 
In order to solve this equation by matrix inversion, it is most convenient to decompose the
matrices $\breve{T}_{s}$, $\breve{V}_s$, and $\breve{G}_{s}$ into a Dirac scalar and a piece 
proportional to $\slashed{p}$ (see also sect.~\ref{sec:bse}).

\section{Loop Integrals}
\label{app:loop_integrals}
\def\theequation{\Alph{section}.\arabic{equation}}
\setcounter{equation}{0}

The tadpole integral is given by
\beqa
I_{\mathrm{M}}^{bj,ai} &=& \int\frac{d^{d}l}{(2\pi)^{d}}\frac{i\delta^{ba}\delta^{ji}}{l^{2}-M_{j}^{2}+i0^{+}} \no \\
&=& \biggl(2M_{j}^{2}\bar\lambda + \frac{M_{j}^{2}}{8\pi^{2}}\ln\biggl(\frac{\mathrm{M}}{\mu}\biggr)\biggr)
    \delta^{ba}\delta^{ji} \,.
\eeqa
Here $\mu$ is the scale of dimensional regularization and 
\begin{displaymath}
\bar\lambda = \frac{\mu^{d-4}}{16\pi^{2}}\biggl(\frac{1}{d-4}-\frac{1}{2}[\ln(4\pi)+\Gamma'(1)+1]\biggr) \,. 
\end{displaymath}
In practice, we will use the ${\overline{MS}}$ renormalization scheme, {\it i.e.}\  terms proportional 
$\bar\lambda$ will be dropped from all expressions.
In the above result, we have neglected terms of $O(4-d)$ since we are interested in 
the limit $d\rightarrow 4$. The diagonal entries in the matrix $I_{\mathrm{M}}$ of eq.~(\ref{eq:IM}) 
are given by the above expression for the tadpole integral, where $M$ is given by the meson 
mass of the respective channel. Moreover, the scale $\mu$ can vary between the different channels.
We shall also define a diagonal matrix $I_{\mathrm{B}}$ with entries given by tadpole integrals where 
$M_{j}$ is replaced by the baryon mass $m_{a}$ of each channel.

In the following, we will leave out the $+i0^{+}$ prescription in the loop integrals 
for brevity. The fundamental scalar loop integral reads
\beqa \label{eq:bubble}
I_{0}(p) &=& \int\frac{d^{d}l}{(2\pi)^{d}}\frac{i}{[(p-l)^{2}-m^{2}][l^{2}-M^{2}]}   \no \\
&=& \frac{1}{16\pi^{2}}\bigg\{-1+\ln\biggl(\frac{m^{2}}{\mu^{2}}\biggr) \no \\
&&+ \frac{M^{2}-m^{2}+p^{2}}{2p^{2}}\ln\biggr(\frac{M^{2}}{m^{2}}\biggl)\no \\ 
&&- \frac{4|\mathbf{q}|}{\sqrt{p^{2}}}
    \artanh\biggl(\frac{2|\mathbf{q}|\sqrt{p^{2}}}{(m+M)^{2}-p^{2}}\biggr)\bigg\} \,,  
\eeqa
where   
\beq \label{eq:qcms}
|\mathbf{q}| = \frac{\sqrt{(p^{2}-(m+M)^{2})(p^{2}-(m-M)^{2})}}{2\sqrt{p^{2}}}
\eeq
is equal to the modulus of the center-of-mass three momentum for a system with particle 
masses $m$ and $M$, and $p^{2}$ the total invariant energy squared of the system.

It is useful to define a diagonal matrix $I_{\textrm{MB}}(p)$, with elements given by the above 
loop integral $I_{0}$ of eq.~(\ref{eq:bubble}), where $m$ is the baryon mass $m_{a}$ 
and $M$ is the meson mass $M_{i}$ of the respective channels. Similarly, we will use 
matrices $I_{\textrm{BB}}$ and $I_{\textrm{MM}}$, where $m$ and $M$ are both given by either the baryon 
mass $m_{a}$ (for $I_{\textrm{BB}}$) or by $M_{i}$ (for $I_{\textrm{MM}}$), respectively.
In the same manner we define 
\beq
I_{\textrm{MBB}}^{bj,ai} = \int\!\frac{d^{d}l}{(2\pi)^{d}}
\frac{i\delta^{ba}\delta^{ji}}{[(p-l)^{2}-m_{a}^{2}][(p_{1}-l)^{2}-m_{a}^{2}][l^{2}-M_{i}^{2}]} .
\eeq
An analogous definition applies for $I_{\textrm{MMB}}$. Only the diagonal elements 
are needed here, since the coupling of the photon does not alter the meson-baryon channel. 
The loop integrals occurring in $I_{\textrm{MBB}}$ and $I_{\textrm{MMB}}$ can be expressed 
in terms of Spence functions \cite{Veltman}.
In the course of the calculation, loop integrals with a tensor structure in the integrand 
are also encountered. For example,
\beq
\int\frac{d^{d}l}{(2\pi)^{d}}\frac{i\delta^{ba}\delta^{ji}l^{\mu}}{[(p-l)^{2}-m_{a}^{2}][l^{2}-M_{i}^{2}]}
= p^{\mu}[I^{(1)}_{\textrm{MB}}]^{bj,ai}
\eeq
with
\beq \label{eq:I1MB}
[I^{(1)}_{\textrm{MB}}]^{bj,ai} 
= \frac{1}{2p^{2}}\biggl[(p^{2}+M_{i}^{2}-m_{a}^{2})I_{\textrm{MB}}^{bj,ai} 
  + I_{\mathrm{B}}^{bj,ai}-I_{\mathrm{M}}^{bj,ai}\biggr].
\eeq
The last expression is simplified for equal masses in the propagators, so that, {\it e.g.}, 
(in matrix notation) 
\beq
\int\frac{d^{d}l}{(2\pi)^{d}}il^{\mu}\Delta(p-l)\Delta(l) = \frac{p^{\mu}}{2}I_{\textrm{MM}}(p) \,.
\eeq
We also need
\beq
\int\frac{d^{d}l}{(2\pi)^{d}}il^{\mu}l^{\nu}\Delta(p-l)\Delta(l) 
= g^{\mu\nu}I_{\textrm{MM}}^{(a)}(p)+\frac{p^{\mu}p^{\nu}}{p^{2}}I_{\textrm{MM}}^{(b)}(p) ,
\eeq
where the coefficients of the tensor structures are given by
\begin{align}
\label{eq:imma} (d-1)I_{\textrm{MM}}^{(a)}(p) 
&= (M^{2}-\frac{1}{4}p^{2})I_{\textrm{MM}}(p)+\frac{1}{2}I_{\mathrm{M}} \,, \\
\label{eq:immb} (d-1)I_{\textrm{MM}}^{(b)}(p) 
&= (\frac{d}{4}p^{2}-M^{2})I_{\textrm{MM}}(p)+(\frac{d}{2}-1)I_{\mathrm{M}} \,.
\end{align}
Here we have employed the meson mass matrix
\begin{displaymath}
M^{bj,ai} = \delta^{ba}\delta^{ji}M_{j}.
\end{displaymath}
defined in analogy to the baryon mass matrix $m$, and the term $p^{2}$ in the last 
two equations is of course to be understood as being multiplied by the identity matrix 
in channel space.

The matrix $G$ of eq.~(\ref{eq:G}) can be given in terms of scalar loop integrals already defined:
\beq \label{eq:Gexplizit}
G(p) = \frac{\slashed{p}}{2p^{2}}\biggl([p^{2}-M^{2}+m^{2}]I_{\textrm{MB}}(p)+I_{\mathrm{M}}-I_{\mathrm{B}}\biggr)
       + mI_{\textrm{MB}}(p).
\eeq
Integrals with a vector or tensor structure in the numerator will also be required in 
the case of three propagators. In particular, two of the three masses in the propagators 
will be equal, as already mentioned above. In the following, we will consider the case 
$\textrm{MBB}$, noting that the other case ($\textrm{MMB}$) is then given simply by 
interchanging $m$ and $M$. First,
\begin{align}
&\int\frac{d^{d}l}{(2\pi)^{d}}
\frac{i\delta^{ba}\delta^{ji}l^{\mu}}{[(p-l)^{2}-m_{a}^{2}][(p_{1}-l)^{2}-m_{a}^{2}][l^{2}-M_{j}^{2}]} \no \\[.5ex]
&= A^{bj,ai}(p_{1},p)(p_{1}+p)^{\mu}+B^{bj,ai}(p_{1},p)(p_{1}-p)^{\mu} \,.
\end{align}
The channel matrices $A$ and $B$ defined by this equation read
\begin{align}
\label{eq:A} A &= \frac{1}{2D}\bigg\{\Bigl(4\bar M^{2}-\frac{\Delta p^{4}}{k^{2}}\Bigr)I_{\textrm{MBB}}
                  +2I_{\textrm{BB}}(k) \no \\
&-\Bigl(1-\frac{\Delta p^{2}}{k^{2}}\Bigr)I_{\textrm{MB}}(p_{1})
 -\Bigl(1+\frac{\Delta p^{2}}{k^{2}}\Bigr)I_{\textrm{MB}}(p)\big\} \,, \\
\label{eq:B} B &= \frac{\Delta p^{2}}{2k^{2}D}\bigg\{(4\bar M^{2}+k^{2}-4\bar p^{2})I_{\textrm{MBB}}
                  +2I_{\textrm{BB}}(k) \no \\
&-\Bigl(1-\frac{4\bar p^{2}-k^{2}}{\Delta p^{2}}\Bigr)I_{\textrm{MB}}(p_{1}) \no \\
&-\Bigl(1+\frac{4\bar p^{2}-k^{2}}{\Delta p^{2}}\Bigr)I_{\textrm{MB}}(p)\bigg\} \,. 
\end{align}
Here we have used the following notation:
\beqa
\bar p^{2} &=& \frac{p_{1}^{2}+p^{2}}{2} ,\\
\bar M^{2} &=& \frac{1}{2}(\bar p^{2}+M^{2}-m^{2}), \\
\Delta p^{2} &=& p^{2}-p_{1}^{2} ,\\
k &=& p - p_{1}, \\
D &=& 4\bar p^{2}-k^{2}-\frac{\Delta p^{4}}{k^{2}}, 
\eeqa
and the abbreviation $\Delta p^{4}\equiv (\Delta p^{2})^{2}$.
The loop integral with three propagators and tensor structure is given by
\begin{align}
I_{\Delta}^{\mu\nu} &= \int\frac{d^{d}l}{(2\pi)^{d}}
\frac{i\delta^{ba}\delta^{ji}l^{\mu}l^{\nu}}{[(p-l)^{2}-m_{a}^{2}][(p_{1}-l)^{2}-m_{a}^{2}][l^{2}-M_{j}^{2}]}\no\\ 
&= C_{1}^{bj,ai}g^{\mu\nu} \no \\
&+ C_{2}^{bj,ai}(p_{1}+p)^{\mu}(p_{1}+p)^{\nu}+C_{3}^{bj,ai}(p_{1}-p)^{\mu}(p_{1}-p)^{\nu} \no \\
&+ C_{4}^{bj,ai}((p_{1}+p)^{\mu}(p_{1}-p)^{\nu}+(p_{1}+p)^{\nu}(p_{1}-p)^{\mu}) \,.
\end{align}
The coefficient matrices $C_{i}$ read as follows:
\begin{align}
C_{1} &= \frac{1}{d-2}\bigg\{M^{2}I_{\textrm{MBB}}+\frac{1}{2}I_{\textrm{BB}}(k)-2\bar M^{2}A
        +\frac{\Delta p^{2}}{2}B\bigg\} \,, \displaybreak[0]\\
C_{2} &= \frac{1}{k^{2}D}\bigg\{k^{2}(M^{2}I_{\textrm{MBB}}+I_{\textrm{BB}}(k)) \no \\
      & +\frac{\Delta p^{2}}{2}(k^{2}B-\Delta p^{2}A)
        -\frac{k^{2}}{4}(I^{(1)}_{\textrm{MB}}(p_{1})+I^{(1)}_{\textrm{MB}}(p))\no \\
      & +\frac{\Delta p^{2}}{4}(I^{(1)}_{\textrm{MB}}(p_{1})-I^{(1)}_{\textrm{MB}}(p))
        -(d-1)k^{2}C_{1}\bigg\} \,, \displaybreak[0]\\
C_{3} &= \frac{1}{k^{2}D}\bigg\{(4\bar p^{2}-k^{2})(M^{2}I_{\textrm{MBB}}
        +\frac{1}{2}I_{\textrm{BB}}(k)) \no \\
      & -2\bar M^{2}((4\bar p^{2}-k^{2})A-\Delta p^{2}B) \no \\
      & +\frac{1}{4}(4\bar p^{2}-\Delta p^{2}-k^{2})I_{\textrm{MB}}^{(1)}(p_{1}) \no \\
      & +\frac{1}{4}(4\bar p^{2}+\Delta p^{2}-k^{2})I_{\textrm{MB}}^{(1)}(p)
        -(d-1)(4\bar p^{2}-k^{2})C_{1}\bigg\} \,, \displaybreak[0]\\
C_{4} &= \frac{1}{k^{2}D}\bigg\{\Delta p^{2}(M^{2}I_{\textrm{MBB}}+I_{\textrm{BB}}(k)) \no \\
      & -\frac{\Delta p^{2}}{2}((4\bar p^{2}-k^{2})A-\Delta p^{2}B) \no \\
      & +\frac{1}{4}(4\bar p^{2}-\Delta p^{2}-k^{2})I_{\textrm{MB}}^{(1)}(p_{1}) \no \\
      & -\frac{1}{4}(4\bar p^{2}+\Delta p^{2}-k^{2})I_{\textrm{MB}}^{(1)}(p)
        -(d-1)\Delta p^{2}C_{1}\bigg\} \,.
\end{align}
$A$ and $B$ are given in eqs.~(\ref{eq:A}) and (\ref{eq:B}), while $I^{(1)}_{\textrm{MB}}$ 
is defined in eq.~(\ref{eq:I1MB}). Note that in the limit $d\rightarrow 4$, $C_{1}$ will 
aquire an additional finite contribution due to the divergent terms in the loop integrals.
In particular,
\beqa
C_{1} &\rightarrow& C_{1}(d=4)-\frac{1}{64\pi^{2}}, \\
(d-1)C_{1} &\rightarrow& 3C_{1}(d=4)+\frac{1}{32\pi^{2}}.
\eeqa
As already mentioned, the results for the $\textrm{MMB}$ case can be obtained from the 
above formulae by interchanging $m \leftrightarrow M$. The corresponding coefficients 
will be denoted by $\tilde A$,$\tilde B$ and $\tilde C_{i}$.  
Finally, the preceding results can be used to derive the decompositions
\begin{align} \label{eq:DefFi}
I_{F} &= \int\frac{d^{d}l}{(2\pi)^{d}}iS(\slashed{p}-\slashed{l})eQ_{\mathrm{B}}
         \gamma^{\mu}S(\slashed{p}_{1}-\slashed{l})\Delta(l) \no \\
&= \gamma^{\mu}F_{1} + \slashed{p}\gamma^{\mu}F_{2}+\gamma^{\mu}\slashed{p_{1}}F_{3}
  +\slashed{p}\gamma^{\mu}\slashed{p_{1}}F_{4}+p^{\mu}F_{5} \no \\
&\quad + p_{1}^{\mu}F_{6}+p^{\mu}\slashed{p}F_{7}+p^{\mu}\slashed{p_{1}}F_{8}
       + p_{1}^{\mu}\slashed{p}F_{9}+p_{1}^{\mu}\slashed{p_{1}}F_{10}
\end{align}
and 
\begin{align} \label{eq:DefFitilde}
I_{\tilde F} &= \int\frac{d^{d}l}{(2\pi)^{d}}i\Delta(p-l)eQ_{\mathrm{M}}(p+p_{1}-2l)^{\mu}
                \Delta(p_{1}-l)S(\slashed{l}) \no \\
&= \gamma^{\mu}\tilde F_{1} + \slashed{p}\gamma^{\mu}\tilde F_{2}+\gamma^{\mu}\slashed{p_{1}}\tilde F_{3}
  +\slashed{p}\gamma^{\mu}\slashed{p_{1}}\tilde F_{4}+p^{\mu}\tilde F_{5}\no \\
&\quad + p_{1}^{\mu}\tilde F_{6}+p^{\mu}\slashed{p}\tilde F_{7}+p^{\mu}\slashed{p_{1}}\tilde F_{8}
       + p_{1}^{\mu}\slashed{p}\tilde F_{9}+p_{1}^{\mu}\slashed{p_{1}}\tilde F_{10} \,, 
\end{align}
where the coefficients of the Lorentz structures are given by
\begin{align*}
F_{1}  &= eQ_{\mathrm{B}}(2C_{1}+(m^{2}-M^{2})I_{\textrm{MBB}}-I_{\textrm{BB}}(k) \\
       &\quad + p_{1}^{2}(A+B)+p^{2}(A-B)) \,, \displaybreak[0]\\
F_{2}  &= eQ_{\mathrm{B}}mI_{\textrm{MBB}} \,, \displaybreak[0]\\
F_{3}  &= F_{2} \,, \displaybreak[0]\\
F_{4}  &= eQ_{\mathrm{B}}(I_{\textrm{MBB}}-2A) \,, \displaybreak[0]\\   
F_{5}  &= 2eQ_{\mathrm{B}}m(B-A) \,, \displaybreak[0]\\
F_{6}  &= -2eQ_{\mathrm{B}}m(B+A) \,, \displaybreak[0]\\
F_{7}  &= 2eQ_{\mathrm{B}}(B-A+C_{2}+C_{3}-2C_{4}) \,, \displaybreak[0]\\
F_{8}  &= 2eQ_{\mathrm{B}}(C_{2}-C_{3}) \,, \displaybreak[0]\\
F_{9}  &= F_{8} \,, \displaybreak[0]\\
F_{10} &= 2eQ_{\mathrm{B}}(C_{2}+C_{3}+2C_{4}-A-B) \,, \displaybreak[0]\\
\tilde F_{1}  &= -2eQ_{\mathrm{M}}\tilde C_{1} \,, \displaybreak[0]\\
\tilde F_{2}  &= \tilde F_{3} = \tilde F_{4} = 0 \,, \displaybreak[0]\\
\tilde F_{5}  &= eQ_{\mathrm{M}}(mI_{\textrm{MMB}}+2m(\tilde B - \tilde A)) \,, \displaybreak[0]\\
\tilde F_{6}  &= eQ_{\mathrm{M}}(mI_{\textrm{MMB}}-2m(\tilde B + \tilde A)) \,, \displaybreak[0]\\
\tilde F_{7}  &= eQ_{\mathrm{M}}(\tilde A - \tilde B -2(\tilde C_{2}+\tilde C_{3}-2\tilde C_{4})) \,, \displaybreak[0]\\
\tilde F_{8}  &= eQ_{\mathrm{M}}(\tilde A + \tilde B -2(\tilde C_{2}-\tilde C_{3})) \,, \displaybreak[0]\\
\tilde F_{9}  &= eQ_{\mathrm{M}}(\tilde A - \tilde B -2(\tilde C_{2}-\tilde C_{3})) \,, \displaybreak[0]\\
\tilde F_{10} &= eQ_{\mathrm{M}}(\tilde A + \tilde B -2(\tilde C_{2}+\tilde C_{3}+2\tilde C_{4})) \,. 
\end{align*}
The following relations are helpful, {\it e.g.}\  when checking gauge invariance:
\begin{align*}
G_{1}(p) &= \bar p^{2}F_{4}-\frac{\Delta p^{2}}{2}(F_{4}+F_{7}+F_{9}) \\
         &\quad +\frac{k^{2}}{2}(F_{9}-F_{7})-F_{1} \\
         &= \frac{k^{2}}{2}(\tilde F_{9}-\tilde F_{7})
           -\frac{\Delta p^{2}}{2}(\tilde F_{9}+\tilde F_{7})-\tilde F_{1} \,, \\
G_{1}(p_{1}) &= \bar p^{2}F_{4}+\frac{\Delta p^{2}}{2}(F_{4}+F_{8}+F_{10}) \\
             &\quad + \frac{k^{2}}{2}(F_{8}-F_{10})-F_{1} \\
             &= \frac{k^{2}}{2}(\tilde F_{8}-\tilde F_{10})
               +\frac{\Delta p^{2}}{2}(\tilde F_{8}+\tilde F_{10})-\tilde F_{1} \,,
\end{align*}
and also
\begin{align*}
G_{0}(p)-G_{0}(p_{1}) &= \frac{k^{2}}{2}(F_{6}-F_{5})-\frac{\Delta p^{2}}{2}(2F_{2}+F_{5}+F_{6}) \\
                      &= \frac{k^{2}}{2}(\tilde F_{6}-\tilde F_{5})
                        -\frac{\Delta p^{2}}{2}(\tilde F_{5}+\tilde F_{6}) \,.
\end{align*}
As a further check of the calculation we have utilized the routines provided
by the FeynCalc package \cite{Mertig:1990an}.

\section{Decomposition of the amplitudes}
\label{app:decomp}
\def\theequation{\Alph{section}.\arabic{equation}}
\setcounter{equation}{0}

We start with the amplitudes of eqs.~(\ref{eq:s})-(\ref{eq:t}).
\beq \label{eq:Ssdecomp}
S_{s}^{\mu} = (\slashed{q}\slashed{p}S_{s}^{\slashed{q}\slashed{p}\gamma} 
+ \slashed{q}S_{s}^{\slashed{q}\gamma}+\slashed{p}S_{s}^{\slashed{p}\gamma}
+S_{s}^{\gamma})\gamma^{\mu}\gamma_{5}
\eeq
with
\begin{eqnarray*}
S_{s}^{\slashed{q}\slashed{p}\gamma} &=& \frac{e}{m_{p}^{2}-s}(\Gamma_{2}(p)-m_{p}\Gamma_{1}(p)) \,, \\
S_{s}^{\slashed{q}\gamma} &=& \frac{e}{m_{p}^{2}-s}(s\Gamma_{1}(p)-m_{p}\Gamma_{2}(p)) \,,\\
S_{s}^{\slashed{p}\gamma} &=& \frac{e}{m_{p}^{2}-s}(\Gamma_{4}(p)-m_{p}\Gamma_{3}(p)) \,,\\
S_{s}^{\gamma} &=& \frac{e}{m_{p}^{2}-s}(s\Gamma_{3}(p)-m_{p}\Gamma_{4}(p)) \,.
\end{eqnarray*} 
Here, $m_{p}$ is the proton mass and $s\equiv p^{2}=(p_{1}+k)^{2}$. Moreover,
\beq \label{eq:Sudecomp}
S_{u}^{\mu} = q^{\mu}\gamma_{5}S_{u}^{q} + \slashed{q}\gamma^{\mu}\gamma_{5}S_{u}^{\slashed{q}\gamma}
              + \gamma^{\mu}\gamma_{5}S_{u}^{\gamma}
\eeq
with
\begin{align*}
S_{u}^{q} &= -2 S_{u}^{\slashed{q}\gamma} = \frac{2eQ_{\mathrm{B}}}{u-m^{2}}Y_{3} \,,\\
S_{u}^{\gamma} &= \frac{eQ_{\mathrm{B}}}{m^{2}-u}\big(\Gamma_{1}(p_{1})m_{p}(u-m_{p}^{2})
                  -\Gamma_{2}(p_{1})(u-m_{p}^{2}\big) \\
               &\quad + (m_{p}^{2}-m_{p}m)\Gamma_{3}(p_{1})+(m-m_{p})\Gamma_{4}(p_{1})) \,,
\end{align*}
and
\begin{align} \label{eq:Stdecomp}
S_{t}^{\mu} &= \big(p^{\mu}S_{t}^{p} + p_{1}^{\mu}S_{t}^{p_{1}}+q^{\mu}S_{t}^{q}
               +\slashed{q}q^{\mu}S_{t}^{\slashed{q}q} + \slashed{p}q^{\mu}S_{t}^{\slashed{p}q} \no \\
            &\quad + \slashed{q}p^{\mu}S_{t}^{\slashed{q}p}+\slashed{p}p^{\mu}S_{t}^{\slashed{p}p}
               +\slashed{q}p_{1}^{\mu}S_{t}^{\slashed{q}p_{1}}
               +\slashed{p}p_{1}^{\mu}S_{t}^{\slashed{p}p_{1}})\gamma_{5} \,,
\end{align}
with
\begin{align*}
S_{t}^{p} &= \frac{eQ_{\mathrm{M}}}{t-M^{2}}Y_{1} \,, &
S_{t}^{p_{1}} &= -S_{t}^{p} \,, &
S_{t}^{q} &= -2S_{t}^{p} \,, \\
S_{t}^{\slashed{q}q} &= -\frac{2eQ_{\mathrm{M}}}{t-M^{2}}Y_{2} \,, &
S_{t}^{\slashed{p}q} &= -S_{t}^{\slashed{q}q} \,, &
S_{t}^{\slashed{q}p} &= \frac{eQ_{\mathrm{M}}}{t-M^{2}}Y_{2} \,, \\
S_{t}^{\slashed{p}p} &= -S_{t}^{\slashed{q}p} \,, &
S_{t}^{\slashed{q}p_{1}} &= -S_{t}^{\slashed{q}p} \,, &
S_{t}^{\slashed{p}p_{1}} &= +S_{t}^{\slashed{q}p} \,. 
\end{align*}
The foregoing results deserve some comments. We use the abbreviations
\begin{align}
\label{eq:Y1} Y_{1} &= m_{p}^{2}\Gamma_{1}(p_{1})-m_{p}(\Gamma_{2}(p_{1})+\Gamma_{3}(p_{1}))+\Gamma_{4}(p_{1}) \,, \\
\label{eq:Y2} Y_{2} &= \Gamma_{2}(p_{1})-m_{p}\Gamma_{1}(p_{1}) \,, \\
\label{eq:Y3} Y_{3} &= (m_{p}^{2}+m_{p}m)\Gamma_{1}(p_{1})-(m+m_{p})\Gamma_{2}(p_{1}) \no \\ 
                    &\quad - m_{p}\Gamma_{3}(p_{1})+\Gamma_{4}(p_{1}) \,. 
\end{align}
Furthermore, we have anticipated that the full amplitude will be multiplied 
by spinors, so we set $\slashed{p}_{1}\gamma_{5}\rightarrow -m_{p}\gamma_{5}$.
The Mandelstam variables $u$ and $t$ are understood to be diagonal matrices in 
channel space and are defined via the center-of-mass (c.m.) relations
\begin{align}
t &= M^{2} + k^{2} - 2 E_{k} E_{q} + 2 |\mathbf{k}| |\mathbf{q}| \cos\theta \,, \\
u &= k^{2} + m_{p}^{2} + m^{2} + M^{2} - s - t \,, 
\end{align}
where $|\mathbf{q}|$ and $|\mathbf{k}|$ are the moduli of the c.m.\ three-momenta 
of the outgoing meson and the incoming photon, respectively (see eq.~(\ref{eq:qcms})),
$\theta$ is the scattering angle in the c.m.\ frame, and the c.m.\ energies are given by
\begin{eqnarray*}
E_{k} &=& \sqrt{|\mathbf{k}|^{2}+k^{2}} \,, \\
E_{q} &=& \sqrt{|\mathbf{q}|^{2}+M^{2}} \,.
\end{eqnarray*}
Of course, multiplication of scalar quantities (such as $k^{2}$ or $m_{p}$) by the 
unit matrix in channel space is implied where necessary 
and fractions involving matrix-valued denominators denote matrix inversions.

Now we come to the graphs where the photon couples to internal meson or baryon lines. 
The decomposition of $S^{\mu}_{\mathrm{B}}$ reads
\begin{align} \label{eq:SBdecomp}
S^{\mu}_{\mathrm{B}} &= \big(\gamma^{\mu}S_{\mathrm{B}}^{\gamma}+p^{\mu}S_{\mathrm{B}}^{p}
+p_{1}^{\mu}S_{\mathrm{B}}^{p_{1}}+\slashed{q}\gamma^{\mu}S_{\mathrm{B}}^{\slashed{q}\gamma} \no \\
&\quad + \slashed{p}\gamma^{\mu}S_{\mathrm{B}}^{\slashed{p}\gamma}
       + \slashed{q}\slashed{p}\gamma^{\mu}S_{\mathrm{B}}^{\slashed{q}\slashed{p}\gamma}
       + \slashed{q}p^{\mu}S_{\mathrm{B}}^{\slashed{q}p}+\slashed{p}p^{\mu}S_{\mathrm{B}}^{\slashed{p}p} \no \\
&\quad + \slashed{q}\slashed{p}p^{\mu}S_{\mathrm{B}}^{\slashed{q}\slashed{p}p}
       +\slashed{q}p_{1}^{\mu}S_{\mathrm{B}}^{\slashed{q}p_{1}}
       +\slashed{p}p_{1}^{\mu}S_{\mathrm{B}}^{\slashed{p}p_{1}}
       +\slashed{q}\slashed{p}p_{1}^{\mu}S_{\mathrm{B}}^{\slashed{q}\slashed{p}p_{1}})\gamma_{5}
\end{align}
with
\begin{align*}
S_{\mathrm{B}}^{\gamma} &= m_{p}[T_{8}F_{3}-T_{5}mF_{3}-T_{5}eQ_{\mathrm{B}}G_{1}(p_{1}) \\
&+ s(T_{3}F_{3}+T_{5}F_{4}+T_{7}F_{4}-T_{3}mF_{4})]Y_{3} \\
&- T_{8}F_{1}Y_{3}+T_{5}mF_{1}Y_{3}+T_{5}eQ_{\mathrm{B}}G_{0}(p_{1})Y_{3} \\
&- T_{5}eQ_{\mathrm{B}}I_{\mathrm{M}}Y_{2} + T_{8}G_{0}(p)eQ_{\mathrm{B}}Y_{2}-T_{5}mG_{0}(p)eQ_{\mathrm{B}}Y_{2} \\
&- s(T_{3}F_{1}Y_{3}+T_{5}F_{2}Y_{3}+T_{7}F_{2}Y_{3}-T_{3}G_{0}(p)eQ_{\mathrm{B}}Y_{2} \\
&- T_{3}mF_{2}Y_{3} - T_{5}G_{1}(p)eQ_{\mathrm{B}}Y_{2} \\
&- T_{7}G_{1}(p)eQ_{\mathrm{B}}Y_{2}+T_{3}mG_{1}(p)eQ_{\mathrm{B}}Y_{2}) \,, \displaybreak[0]\\
S_{\mathrm{B}}^{p} &= m_{p}[T_{8}F_{8}Y_{3}-T_{5}mF_{8}Y_{3}+sT_{3}F_{8}Y_{3}] \\
&- T_{8}F_{5}Y_{3} + T_{5}mF_{5}Y_{3} \\
&- s(T_{3}F_{5}Y_{3}+T_{5}F_{7}Y_{3}+T_{7}F_{7}Y_{3}-T_{3}mF_{7}Y_{3}) \,, \displaybreak[0]\\
S_{\mathrm{B}}^{p_{1}} &= m_{p}[T_{8}F_{10}Y_{3}-T_{5}mF_{10}Y_{3}+sT_{3}F_{10}Y_{3}]-T_{8}F_{6}Y_{3} \\
&+ T_{5}mF_{6}Y_{3}-s(T_{3}F_{6}+T_{5}F_{9}+T_{7}F_{9}-T_{3}mF_{9})Y_{3} \,, \displaybreak[0]\\
S_{\mathrm{B}}^{\slashed{q}\gamma} &= m_{p}[T_{6}F_{3}-T_{2}mF_{3}-T_{2}eQ_{\mathrm{B}}G_{1}(p_{1}) \\ 
&+ s(T_{1}F_{3}+T_{2}F_{4}+T_{4}F_{4} - T_{1}mF_{4})]Y_{3} \\ 
&- T_{6}F_{1}Y_{3}+T_{2}mF_{1}Y_{3}+T_{2}eQ_{\mathrm{B}}G_{0}(p_{1})Y_{3} \\
&- T_{2}eQ_{\mathrm{B}}I_{\mathrm{M}}Y_{2} + T_{6}G_{0}(p)eQ_{\mathrm{B}}Y_{2}-T_{2}mG_{0}(p)eQ_{\mathrm{B}}Y_{2} \\ 
&- s(T_{1}F_{1}Y_{3}+T_{2}F_{2}Y_{3}+T_{4}F_{2}Y_{3}-T_{1}G_{0}(p)eQ_{\mathrm{B}}Y_{2} \\ 
&- T_{1}mF_{2}Y_{3}-T_{2}G_{1}(p)eQ_{\mathrm{B}}Y_{2}-T_{4}G_{1}(p)eQ_{\mathrm{B}}Y_{2} \\ 
&+ T_{1}mG_{1}(p)eQ_{\mathrm{B}}Y_{2}) \,, \displaybreak[0]\\
S_{\mathrm{B}}^{\slashed{p}\gamma} &= m_{p}[T_{5}F_{3}+T_{7}F_{3}+T_{8}F_{4}-T_{3}mF_{3} \\
&- T_{3}eQ_{\mathrm{B}}G_{1}(p_{1}) - T_{5}mF_{4} + sT_{3}F_{4}]Y_{3} \\
&- T_{5}F_{1}Y_{3}-T_{7}F_{1}Y_{3}-T_{8}F_{2}Y_{3}+T_{3}mF_{1}Y_{3} \\ 
&+ T_{3}eQ_{\mathrm{B}}G_{0}(p_{1})Y_{3}-T_{3}eQ_{\mathrm{B}}I_{\mathrm{M}}Y_{2} \\ 
&+ T_{5}G_{0}(p)eQ_{\mathrm{B}}Y_{2} + T_{5}mF_{2}Y_{3} + T_{7}G_{0}(p)eQ_{\mathrm{B}}Y_{2} \\ 
&+ T_{8}G_{1}(p)eQ_{\mathrm{B}}Y_{2}-T_{3}mG_{0}(p)eQ_{\mathrm{B}}Y_{2} \\ 
&- T_{5}mG_{1}(p)eQ_{\mathrm{B}}Y_{2}+s(T_{3}G_{1}(p)eQ_{\mathrm{B}}Y_{2}-T_{3}F_{2}Y_{3}) \,, \displaybreak[0]\\
S_{\mathrm{B}}^{\slashed{q}\slashed{p}\gamma} &= m_{p}[T_{2}F_{3}+T_{4}F_{3}+T_{6}F_{4}-T_{1}mF_{3} \\
&- T_{1}eQ_{\mathrm{B}}G_{1}(p_{1}) - T_{2}mF_{4}+sT_{1}F_{4}]Y_{3} \\
&- T_{2}F_{1}Y_{3}-T_{4}F_{1}Y_{3}-T_{6}F_{2}Y_{3} \\
&+ T_{1}mF_{1}Y_{3} + T_{1}eQ_{\mathrm{B}}G_{0}(p_{1})Y_{3}-T_{1}eQ_{\mathrm{B}}I_{\mathrm{M}}Y_{2} \\
&+ T_{2}G_{0}(p)eQ_{\mathrm{B}}Y_{2} + T_{2}mF_{2}Y_{3}+T_{4}G_{0}(p)eQ_{\mathrm{B}}Y_{2} \\
&+ T_{6}G_{1}(p)eQ_{\mathrm{B}}Y_{2}-T_{1}mG_{0}(p)eQ_{\mathrm{B}}Y_{2} \\
&- T_{2}mG_{1}(p)eQ_{\mathrm{B}}Y_{2} - s(T_{1}F_{2}Y_{3}-T_{1}G_{1}(p)eQ_{\mathrm{B}}Y_{2}) \,, \displaybreak[0]\\
S_{\mathrm{B}}^{\slashed{q}p} &= m_{p}[T_{6}F_{8}-T_{2}mF_{8}+sT_{1}F_{8}]Y_{3} \\
&- T_{6}F_{5}Y_{3}+T_{2}mF_{5}Y_{3} \\
&- s(T_{1}F_{5}+T_{2}F_{7}+T_{4}F_{7}-T_{1}mF_{7})Y_{3} \,, \displaybreak[0]\\
S_{\mathrm{B}}^{\slashed{p}p} &= m_{p}[T_{5}F_{8}+T_{7}F_{8}-T_{3}mF_{8}]Y_{3} \\ 
&- (T_{5}F_{5}+T_{7}F_{5}+T_{8}F_{7}-T_{3}mF_{5} \\
&- T_{5}mF_{7}+sT_{3}F_{7})Y_{3} \,, \displaybreak[0]\\
S_{\mathrm{B}}^{\slashed{q}\slashed{p}p} &= m_{p}[T_{2}F_{8}+T_{4}F_{8}-T_{1}mF_{8}]Y_{3} \\
&- (T_{2}F_{5}+T_{4}F_{5} + T_{6}F_{7}-T_{1}mF_{5} \\
&- T_{2}mF_{7}+sT_{1}F_{7})Y_{3} \,, \displaybreak[0]\\
S_{\mathrm{B}}^{\slashed{q}p_{1}} &= m_{p}[T_{6}F_{10}-T_{2}mF_{10}+sT_{1}F_{10}]Y_{3} \\
&- T_{6}F_{6}Y_{3}+T_{2}mF_{6}Y_{3} \\
&- s(T_{1}F_{6}+T_{2}F_{9}+T_{4}F_{9}-T_{1}mF_{9})Y_{3} \,, \displaybreak[0]\\
S_{\mathrm{B}}^{\slashed{p}p_{1}} &= m_{p}[T_{5}F_{10}+T_{7}F_{10}-T_{3}mF_{10}]Y_{3} \\
&- (T_{5}F_{6}+T_{7}F_{6}+T_{8}F_{9} - T_{3}mF_{6} \\
&- T_{5}mF_{9}+sT_{3}F_{9})Y_{3} \,, \displaybreak[0]\\
S_{\mathrm{B}}^{\slashed{q}\slashed{p}p_{1}} &= m_{p}[T_{2}F_{10}+T_{4}F_{10}-T_{1}mF_{10}]Y_{3} \\
&- (T_{2}F_{6}+T_{4}F_{6} + T_{6}F_{9}-T_{1}mF_{6} \\
&- T_{2}mF_{9}+sT_{1}F_{9})Y_{3} \,,
\end{align*}
while the decomposition of $S_{\mathrm{M}}^{\mu}$ reads
\begin{align} \label{eq:SMdecomp}
S^{\mu}_{\mathrm{M}} &= \big(\gamma^{\mu}S_{\mathrm{M}}^{\gamma}+p^{\mu}S_{\mathrm{M}}^{p}
+p_{1}^{\mu}S_{\mathrm{M}}^{p_{1}}+\slashed{q}\gamma^{\mu}S_{\mathrm{M}}^{\slashed{q}\gamma}
       + \slashed{p}\gamma^{\mu}S_{\mathrm{M}}^{\slashed{p}\gamma} \no \\
&\quad + \slashed{q}\slashed{p}\gamma^{\mu}S_{\mathrm{M}}^{\slashed{q}\slashed{p}\gamma}
       + \slashed{q}p^{\mu}S_{\mathrm{M}}^{\slashed{q}p}+\slashed{p}p^{\mu}S_{\mathrm{M}}^{\slashed{p}p}
       + \slashed{q}\slashed{p}p^{\mu}S_{\mathrm{M}}^{\slashed{q}\slashed{p}p} \no \\
&\quad + \slashed{q}p_{1}^{\mu}S_{\mathrm{M}}^{\slashed{q}p_{1}}
       + \slashed{p}p_{1}^{\mu}S_{\mathrm{M}}^{\slashed{p}p_{1}}
       + \slashed{q}\slashed{p}p_{1}^{\mu}S_{\mathrm{M}}^{\slashed{q}\slashed{p}p_{1}}\big)\gamma_{5} \,,
\end{align}
with
\begin{align*}
S_{\mathrm{M}}^{\gamma} &= T_{5}m\tilde F_{1}Y_{3}-T_{8}\tilde F_{1}Y_{3}
  -T_{5}eQ_{\mathrm{M}}d_{1}Y_{2}-sT_{3}\tilde F_{1}Y_{3} \,, \displaybreak[0]\\
S_{\mathrm{M}}^{p} &= m_{p}[T_{8}\tilde F_{8}Y_{3}-T_{5}m\tilde F_{8}Y_{3}
  -T_{5}eQ_{\mathrm{M}}d_{2}Y_{2}+sT_{3}\tilde F_{8}Y_{3}] \\
&- T_{8}\tilde F_{5}Y_{3}+T_{5}m\tilde F_{5}Y_{3}-s(T_{3}\tilde F_{5}Y_{3}+T_{5}\tilde F_{7}Y_{3} \\
&+ T_{7}\tilde F_{7}Y_{3} - T_{3}m\tilde F_{7}Y_{3}+T_{3}eQ_{\mathrm{M}}d_{2}Y_{2}) \,, \displaybreak[0]\\
S_{\mathrm{M}}^{p_{1}} &= m_{p}[T_{8}\tilde F_{10}Y_{3}-T_{5}m\tilde F_{10}Y_{3}+T_{5}eQ_{\mathrm{M}}d_{2}Y_{2} \\
&+ sT_{3}\tilde F_{10}Y_{3}] - T_{8}\tilde F_{6}Y_{3}+T_{5}m\tilde F_{6}Y_{3} \\
&- s(T_{3}\tilde F_{6}+T_{5}\tilde F_{9}+T_{7}\tilde F_{9}-T_{3}m\tilde F_{9})Y_{3} \\
&+ sT_{3}eQ_{\mathrm{M}}d_{2}Y_{2} \,, \displaybreak[0]\\
S_{\mathrm{M}}^{\slashed{q}\gamma} &= T_{2}m\tilde F_{1}Y_{3}-T_{6}\tilde F_{1}Y_{3}
  -sT_{1}\tilde F_{1}Y_{3}-T_{2}eQ_{\mathrm{M}}d_{1}Y_{2} \,, \displaybreak[0]\\
S_{\mathrm{M}}^{\slashed{p}\gamma} &= T_{3}m\tilde F_{1}Y_{3}-T_{5}\tilde F_{1}Y_{3}
  -T_{7}\tilde F_{1}Y_{3}-T_{3}eQ_{\mathrm{M}}d_{1}Y_{2} \,, \displaybreak[0]\\
S_{\mathrm{M}}^{\slashed{q}\slashed{p}\gamma} &= T_{1}m\tilde F_{1}Y_{3}-T_{2}\tilde F_{1}Y_{3}
  -T_{4}\tilde F_{1}Y_{3}-T_{1}eQ_{\mathrm{M}}d_{1}Y_{2} \,, \displaybreak[0]\\
S_{\mathrm{M}}^{\slashed{q}p} &= m_{p}[T_{6}\tilde F_{8}Y_{3}-T_{2}m\tilde F_{8}Y_{3}+sT_{1}\tilde F_{8}Y_{3}\\
&- T_{2}eQ_{\mathrm{M}}d_{2}Y_{2}] - T_{6}\tilde F_{5}Y_{3}+T_{2}m\tilde F_{5}Y_{3} \\
&- s(T_{1}\tilde F_{5}+T_{2}\tilde F_{7}+T_{4}\tilde F_{7}-T_{1}m\tilde F_{7})Y_{3}\\ 
&- sT_{1}eQ_{\mathrm{M}}d_{2}Y_{2} \,, \displaybreak[0]\\
S_{\mathrm{M}}^{\slashed{p}p} &= m_{p}[T_{5}\tilde F_{8}Y_{3}+T_{7}\tilde F_{8}Y_{3}
  -T_{3}m\tilde F_{8}Y_{3}-T_{3}eQ_{\mathrm{M}}d_{2}Y_{2}] \\
&- (T_{5}\tilde F_{5}+T_{7}\tilde F_{5}+T_{8}\tilde F_{7}-T_{3}m\tilde F_{5}-T_{5}m\tilde F_{7})Y_{3} \\
&- sT_{3}\tilde F_{7}Y_{3}-T_{5}eQ_{\mathrm{M}}d_{2}Y_{2} \,, \displaybreak[0]\\
S_{\mathrm{M}}^{\slashed{q}\slashed{p}p} &= m_{p}[T_{2}\tilde F_{8}Y_{3}+T_{4}\tilde F_{8}Y_{3}
  -T_{1}m\tilde F_{8}Y_{3}-T_{1}eQ_{\mathrm{M}}d_{2}Y_{2}] \\
&- (T_{2}\tilde F_{5}+T_{4}\tilde F_{5}+T_{6}\tilde F_{7}-T_{1}m\tilde F_{5}-T_{2}m\tilde F_{7}) \\
&- sT_{1}\tilde F_{7}Y_{3}-T_{2}eQ_{\mathrm{M}}d_{2}Y_{2} ,\\
S_{\mathrm{M}}^{\slashed{q}p_{1}} &= m_{p}[T_{6}\tilde F_{10}Y_{3}-T_{2}m\tilde F_{10}Y_{3}
  +sT_{1}\tilde F_{10}Y_{3} \\
&+ T_{2}eQ_{\mathrm{M}}d_{2}Y_{2}] - T_{6}\tilde F_{6}Y_{3}+T_{2}m\tilde F_{6}Y_{3} \\
&+ sT_{1}eQ_{\mathrm{M}}d_{2}Y_{2} \\
&- s(T_{1}\tilde F_{6}+T_{2}\tilde F_{9}+T_{4}\tilde F_{9}-T_{1}m\tilde F_{9})Y_{3} \,, \displaybreak[0]\\
S_{\mathrm{M}}^{\slashed{p}p_{1}} &= m_{p}[T_{5}\tilde F_{10}Y_{3}+T_{7}\tilde F_{10}Y_{3} \\
&- T_{3}m\tilde F_{10}Y_{3}+T_{3}eQ_{\mathrm{M}}d_{2}Y_{2}] \\
&- (T_{5}\tilde F_{6}+T_{7}\tilde F_{6}+T_{8}\tilde F_{9}-T_{3}m\tilde F_{6}-T_{5}m\tilde F_{9})Y_{3} \\
&- sT_{3}\tilde F_{9}Y_{3}+T_{5}eQ_{\mathrm{M}}d_{2}Y_{2} \,, \displaybreak[0]\\
S_{\mathrm{M}}^{\slashed{q}\slashed{p}p_{1}} &= m_{p}[T_{2}\tilde F_{10}Y_{3}+T_{4}\tilde F_{10}Y_{3} \\
&- T_{1}m\tilde F_{10}Y_{3}+T_{1}eQ_{\mathrm{M}}d_{2}Y_{2}] \\
&- (T_{2}\tilde F_{6}+T_{4}\tilde F_{6}+T_{6}\tilde F_{9}-T_{1}m\tilde F_{6}-T_{2}m\tilde F_{9})Y_{3} \\
&- sT_{1}\tilde F_{9}Y_{3}+T_{2}eQ_{\mathrm{M}}d_{2}Y_{2} \,.
\end{align*}
Here we have used the abbreviations $T_{i}\equiv T_{i}(p)$ and
\begin{eqnarray*}
d_{1} &=& -2\biggl[\frac{1}{3}\biggl((M^{2}-\frac{1}{4}k^{2})I_{\textrm{MM}}(k)
          +\frac{1}{2}I_{\mathrm{M}}\biggr)+\delta\biggr] ,\\
d_{2} &=& \frac{1}{2}I_{\textrm{MM}}(k) - \frac{2}{k^{2}}
          \biggl[\frac{1}{3}\biggl((k^{2}-M^{2})I_{\textrm{MM}}(k)+I_{\mathrm{M}}\biggr)-\delta\biggr],\\
\delta &=& \frac{1}{48\pi^{2}}\biggl[\frac{1}{6}k^{2}-M^{2}\biggr].
\end{eqnarray*}
The $T_{i}(p)$ are defined in eq.~(\ref{eq:Tdecomp}), while the $F_{i}$ and $\tilde F_{i}$ 
are defined in eq.~(\ref{eq:DefFi}) and (\ref{eq:DefFitilde}), respectively.
The term $\delta$ stems from the limit $d\rightarrow 4$ in eqs.~(\ref{eq:imma}) and (\ref{eq:immb}).

Next we turn to the class of diagrams derived from the `Kroll-Ruderman'-term 
(called `Class~4' in sect.~\ref{sec:photo}):
\beq \label{eq:SKRdecomp}
S_{\textrm{KR}}^{\mu} = (S_{\textrm{KR}}^{\gamma}+\slashed{q}S_{\textrm{KR}}^{\slashed{q}\gamma}
                       +\slashed{p}S_{\textrm{KR}}^{\slashed{p}\gamma}
                       +\slashed{q}\slashed{p}S_{\textrm{KR}}^{\slashed{q}\slashed{p}\gamma})\gamma^{\mu}\gamma_{5}
\eeq
with
\begin{align*}
S_{\textrm{KR}}^{\gamma} &= eQ_{\mathrm{M}}\hat g + [(T_{8}-T_{5}m+sT_{3})G_{0}(p) \\
&+ s(T_{5}+T_{7}-T_{3}m)G_{1}(p)-T_{5}I_{\mathrm{M}}]eQ_{\mathrm{M}}\hat g \,, \displaybreak[0]\\
S_{\textrm{KR}}^{\slashed{q}\gamma} &= [(T_{6}-T_{2}m+sT_{1})G_{0}(p) \\
&+ s(T_{2}+T_{4}-T_{1}m)G_{1}(p)-T_{2}I_{\mathrm{M}}]eQ_{\mathrm{M}}\hat g \,, \displaybreak[0]\\
S_{\textrm{KR}}^{\slashed{p}\gamma} &= [(T_{5}+T_{7}-T_{3}m)G_{0}(p) \\
&+ (sT_{3}+T_{8}-T_{5}m)G_{1}(p)-T_{3}I_{\mathrm{M}}]eQ_{\mathrm{M}}\hat g \,, \displaybreak[0]\\
S_{\textrm{KR}}^{\slashed{q}\slashed{p}\gamma} &= [(T_{2}+T_{4}-T_{1}m)G_{0}(p)\\
&+ (sT_{1}+T_{6}-T_{2}m)G_{1}(p)-T_{1}I_{\mathrm{M}}]eQ_{\mathrm{M}}\hat g \,.
\end{align*}
Here, the first term in the first line is the contribution from the tree graph, 
and $T_{i}\equiv T_{i}(p)$. The last class of graphs is $S_{WT1}^{\mu}+S_{WT2}^{\mu}$, where
\beq \label{eq:SWT1decomp}
S_{WT1}^{\mu} = \gamma^{\mu}\gamma_{5}S_{WT1}^{\gamma}
\eeq
with
\begin{displaymath}
S_{WT1}^{\gamma} = (Q_{\mathrm{M}}g+gQ_{\mathrm{M}})((G_{0}(p_{1})-m_{p}G_{1}(p_{1}))Y_{3}-I_{\mathrm{M}}Y_{2}) \,,
\end{displaymath}
and
\beq \label{eq:SWT2decomp}
S_{WT2}^{\mu} = (S_{WT2}^{\gamma}+\slashed{q}S_{WT2}^{\slashed{q}\gamma}
               +\slashed{p}S_{WT2}^{\slashed{p}\gamma}
               +\slashed{q}\slashed{p}S_{WT2}^{\slashed{q}\slashed{p}\gamma})\gamma^{\mu}\gamma_{5} 
\eeq
with
\begin{align*}
S_{WT2}^{\gamma} &= [s(T_{3}G_{0}(p)+(T_{5}+T_{7}-T_{3}m)G_{1}(p))\\
&+ (T_{8}-T_{5}m)G_{0}(p)-T_{5}I_{\mathrm{M}}]S_{WT1}^{\gamma} \,, \displaybreak[0]\\
S_{WT2}^{\slashed{q}\gamma} &= [s(T_{1}G_{0}(p)+(T_{2}+T_{4}-T_{1}m)G_{1}(p))\\
&+ (T_{6}-T_{2}m)G_{0}(p)-T_{2}I_{\mathrm{M}}]S_{WT1}^{\gamma} \,, \displaybreak[0]\\
S_{WT2}^{\slashed{p}\gamma} &= [(sT_{3}-T_{5}m+T_{8})G_{1}(p)\\
&+ (T_{5}+T_{7}-T_{3}m)G_{0}(p)-T_{3}I_{\mathrm{M}}]S_{WT1}^{\gamma} \,, \displaybreak[0]\\
S_{WT2}^{\slashed{q}\slashed{p}\gamma} &= [(sT_{1}-T_{2}m+T_{6})G_{1}(p)\\
&+ (T_{2}+T_{4}-T_{1}m)G_{0}(p)-T_{1}I_{\mathrm{M}}]S_{WT1}^{\gamma} \,.
\end{align*}
Adding the contributions of eqs.~(\ref{eq:Ssdecomp})-(\ref{eq:Stdecomp}) and 
eqs.~(\ref{eq:SBdecomp})-(\ref{eq:SWT2decomp}), we can decompose the full 
photoproduction amplitude $\mathcal{M}^{\mu}$ (see eq.~(\ref{eq:Msum})) as follows:
\begin{align} \label{eq:Mmudecomp}
\mathcal{M}^{\mu} &= \gamma^{\mu}\gamma_{5}\mathcal{M}_{1}+q^{\mu}\gamma_{5}\mathcal{M}_{2}
 + p^{\mu}\gamma_{5}\mathcal{M}_{3}+p_{1}^{\mu}\gamma_{5}\mathcal{M}_{4} \no \\
&+ \slashed{q}\gamma^{\mu}\gamma_{5}\mathcal{M}_{5}+\slashed{p}\gamma^{\mu}\gamma_{5}\mathcal{M}_{6}
 + \slashed{q}\slashed{p}\gamma^{\mu}\gamma_{5}\mathcal{M}_{7} \no \\
&+ \slashed{q}q^{\mu}\gamma_{5}\mathcal{M}_{8}+\slashed{p}q^{\mu}\gamma_{5}\mathcal{M}_{9}
 + \slashed{q}p^{\mu}\gamma_{5}\mathcal{M}_{10} \no \\
&+ \slashed{p}p^{\mu}\gamma_{5}\mathcal{M}_{11} + \slashed{q}\slashed{p}p^{\mu}\gamma_{5}\mathcal{M}_{12}
 + \slashed{q}p_{1}^{\mu}\gamma_{5}\mathcal{M}_{13} \no \\
&+ \slashed{p}p_{1}^{\mu}\gamma_{5}\mathcal{M}_{14}+\slashed{q}\slashed{p}p_{1}^{\mu}\gamma_{5}\mathcal{M}_{15} \,.
\end{align}
This can be simplified, using the Dirac equation and momentum conservation, to arrive 
at the operator basis given by the $\mathcal{N}_{k}^{\mu}$ commonly used for photoproduction processes, 
see eq.~(\ref{eq:TfiB}). The relation between the corresponding coefficients $B_{k}$ and the 
functions $\mathcal{M}_{j}$ used in eq.~(\ref{eq:Mmudecomp}) is given in eq.~(\ref{eq:MtoB}).
The decomposition of the amplitudes into the various Dirac structures
is obtained by employing the FeynCalc package \cite{Mertig:1990an}.

\section{Invariant amplitudes}
\label{app:photo_formalism}
\def\theequation{\Alph{section}.\arabic{equation}}
\setcounter{equation}{0}

We consider the reaction 
\begin{displaymath}
\gamma(k) + p(p_{1},m_{1}) \rightarrow B(p_{2},m_{2}) + M(q,M_{\phi})
\end{displaymath}
and define the Mandelstam variables as usual,
\begin{eqnarray*}
s &=& (p_{1}+k)^{2} \,,\\
u &=& (p_{1}-q)^{2} \,,\\
t &=& (p_{2}-p_{1})^{2} \,.
\end{eqnarray*}
They obey the constraint
\begin{displaymath}
s+t+u = m_{1}^{2}+m_{2}^{2}+M_{\phi}^{2}+k^{2} \,.
\end{displaymath}
The amplitude can be decomposed as
\begin{equation} \label{eq:TfiB}
T_{fi}=i\epsilon_{\mu}\bar u_{2}\sum_{k=1}^{8}B_{k}\mathcal{N}_{k}^{\mu}u_{1} \,,
\end{equation}
where the operator basis is given by
\begin{align*}
 \mathcal{N}_{1}^{\mu} &= \gamma_{5}\gamma^{\mu}k \,, 
&\mathcal{N}_{2}^{\mu} &= 2\gamma_{5}P^{\mu} \,,
&\mathcal{N}_{3}^{\mu} &= 2\gamma_{5}q^{\mu} \,, \\
 \mathcal{N}_{4}^{\mu} &= 2\gamma_{5}k^{\mu} \,, 
&\mathcal{N}_{5}^{\mu} &= \gamma_{5}\gamma^{\mu} \,,
&\mathcal{N}_{6}^{\mu} &= \gamma_{5}\slashed{k}P^{\mu} \,, \\
 \mathcal{N}_{7}^{\mu} &= \gamma_{5}\slashed{k}k^{\mu} \,, 
&\mathcal{N}_{8}^{\mu} &= \gamma_{5}\slashed{k}q^{\mu} \,.
\end{align*}
Here, $P = \frac{1}{2}(p_{1}+p_{2})$. The relation to the coefficient functions 
$\mathcal{M}_{j}$ used in eq.~(\ref{eq:Mmudecomp}) is 
\begin{align} \label{eq:MtoB}
B_{1} &= -\mathcal{M}_{5}-\mathcal{M}_{6}+m_{2}\mathcal{M}_{7} \,, \no \displaybreak[0]\\
B_{2} &= \frac{1}{2}\mathcal{M}_{3}+\frac{1}{2}\mathcal{M}_{4}+\mathcal{M}_{5}
  +\mathcal{M}_{6}-m_{2}\mathcal{M}_{7} \no \\
&- \frac{1}{2}(m_{1}+m_{2})\mathcal{M}_{10}-\frac{1}{2}m_{1}\mathcal{M}_{11}
  +\frac{1}{2}(s+m_{1}m_{2})\mathcal{M}_{12} \no \\
&- \frac{1}{2}(m_{1}+m_{2})\mathcal{M}_{13} - \frac{1}{2}m_{1}\mathcal{M}_{14}
  +\frac{1}{2}(s+m_{1}m_{2})\mathcal{M}_{15} \,, \no \displaybreak[0]\\
B_{3} &= \frac{1}{2}\mathcal{M}_{2}+\frac{1}{4}\mathcal{M}_{3}+\frac{1}{4}\mathcal{M}_{4}
  +\frac{1}{2}\mathcal{M}_{5}+\frac{1}{2}\mathcal{M}_{6} \no \\
&- \frac{1}{2}m_{2}\mathcal{M}_{7}-\frac{1}{2}(m_{1}+m_{2})\mathcal{M}_{8}
  -\frac{1}{2}m_{1}\mathcal{M}_{9} \no \\
&- \frac{1}{4}(m_{1}+m_{2})\mathcal{M}_{10}-\frac{1}{4}m_{1}\mathcal{M}_{11}
  +\frac{1}{4}(s+m_{1}m_{2})\mathcal{M}_{12} \no \\
&- \frac{1}{4}(m_{1}+m_{2})\mathcal{M}_{13}-\frac{1}{4}m_{1}\mathcal{M}_{14}
  +\frac{1}{4}(s+m_{1}m_{2})\mathcal{M}_{15} \,, \no \displaybreak[0]\\
B_{4} &= \frac{1}{4}\mathcal{M}_{3}-\frac{1}{4}\mathcal{M}_{4}+\frac{1}{2}\mathcal{M}_{5}
  +\frac{1}{2}\mathcal{M}_{6}-\frac{1}{2}m_{2}\mathcal{M}_{7} \no \\
&- \frac{1}{4}(m_{1}+m_{2})\mathcal{M}_{10}-\frac{1}{4}m_{1}\mathcal{M}_{11}
  +\frac{1}{4}(s+m_{1}m_{2})\mathcal{M}_{12} \no \\
&+ \frac{1}{4}(m_{1}+m_{2})\mathcal{M}_{13}+\frac{1}{4}m_{1}\mathcal{M}_{14}
  -\frac{1}{4}(s+m_{1}m_{2})\mathcal{M}_{15} \,, \no \displaybreak[0]\\
B_{5} &= -\mathcal{M}_{1}+(m_{2}-m_{1})\mathcal{M}_{5}-m_{1}\mathcal{M}_{6} \no \\ 
&- (s-m_{1}m_{2})\mathcal{M}_{7} \,, \no \displaybreak[0]\\
B_{6} &= -\mathcal{M}_{10}-\mathcal{M}_{11}+m_{2}\mathcal{M}_{12} \no\\
&- \mathcal{M}_{13}-\mathcal{M}_{14}+m_{2}\mathcal{M}_{15} \,, \no \displaybreak[0]\\
B_{7} &= -\frac{1}{2}\mathcal{M}_{10}-\frac{1}{2}\mathcal{M}_{11}+\frac{1}{2}m_{2}\mathcal{M}_{12} \no\\
&+ \frac{1}{2}\mathcal{M}_{13}+\frac{1}{2}\mathcal{M}_{14}
  -\frac{1}{2}m_{2}\mathcal{M}_{15} \,, \no \displaybreak[0]\\
B_{8} &= -\mathcal{M}_{8}-\mathcal{M}_{9}-\frac{1}{2}\mathcal{M}_{10}
  -\frac{1}{2}\mathcal{M}_{11}+\frac{1}{2}m_{2}\mathcal{M}_{12} \no \\
&- \frac{1}{2}\mathcal{M}_{13}-\frac{1}{2}\mathcal{M}_{14}+\frac{1}{2}m_{2}\mathcal{M}_{15} \,.
\end{align}

For a gauge invariant amplitude, the following relations for the $B_{i}$ hold:
\begin{eqnarray*}
k^{2}B_{1}+2(k\cdot P)B_{2}+2(k\cdot q)B_{3}+2k^{2}B_{4} &=& 0 , \\
B_{5}+(k\cdot P)B_{6}+k^{2}B_{7}+(k\cdot q)B_{8} &=& 0 .
\end{eqnarray*}
Given these relations, one can eliminate two of the $B_{i}$ (conventionally, one takes 
$B_{3}$ and $B_{5}$) and rewrite the amplitude in a manifestly gauge invariant form:
\begin{equation} \label{eq:TfiA}
T_{fi} = i\bar u_{2}\sum_{i=1}^{6}A_{i}M_{i}u_{1}
\end{equation}
with the operator structures
\begin{align*}
M_{1} &= \frac{1}{2}\gamma_{5}\gamma_{\mu}\gamma_{\nu}F^{\mu\nu} \,, \displaybreak[0]\\
M_{2} &= 2\gamma_{5}P_{\mu}(q-\frac{1}{2}k)_{\nu}F^{\mu\nu} \,, \displaybreak[0]\\ 
M_{3} &= \gamma_{5}\gamma_{\mu}q_{\nu}F^{\mu\nu} \,, \displaybreak[0]\\
M_{4} &= 2\gamma_{5}\gamma_{\mu}P_{\nu}F^{\mu\nu}-(m_{1}+m_{2})M_{1} \,, \displaybreak[0]\\ 
M_{5} &= \gamma_{5}k_{\mu}q_{\nu}F^{\mu\nu} \,, \displaybreak[0]\\
M_{6} &= \gamma_{5}k_{\mu}\gamma_{\nu}F^{\mu\nu} \,,
\end{align*}
where $F^{\mu\nu} = \epsilon^{\mu}k^{\nu}-\epsilon^{\nu}k^{\mu}$. 
The particular form of $M_{4}$ has been chosen such that 
\begin{displaymath}
i\bar u_{2}M_{4}u_{1}\rightarrow 2\sqrt{m_{1}m_{2}}\,
\mathbf{q}\cdot(\mathbf{k}\times\mbox{\boldmath$\epsilon$})
\end{displaymath}
in the nonrelativistic limit, where the baryon masses are large compared to the meson masses 
and three-momenta.

The $A_{i}$ are related to the $B_{i}$ via
\begin{align*}
A_{1} &= B_{1}-\frac{1}{2}(m_{1}+m_{2})B_{6} \,, \displaybreak[0]\\
A_{2} &= \frac{2}{M_{\phi}^{2}-t}B_{2} \,, \displaybreak[0]\\
A_{3} &= -B_{8} \,, \displaybreak[0]\\
A_{4} &= -\frac{1}{2}B_{6} \,, \displaybreak[0]\\
A_{5} &= \frac{2}{s+u-m_{1}^{2}-m_{2}^{2}} \\
&\quad\times \biggl(B_{1} - \frac{s-u+m_{2}^{2}-m_{1}^{2}}{2(M_{\phi}^{2}-t)}B_{2}+2B_{4}\biggr) \,, \displaybreak[0]\\
A_{6} &= B_{7} \,. 
\end{align*}
Following the conventions of Chew, Goldberger, Low and Nambu (CGLN) \cite{CGLN}, 
and Berends, Donnachie and Weaver \cite{Donnachie}, we rewrite this once more, 
making use of the standard representation of spinors and Gamma matrices.  
In terms of Pauli spinors and matrices, one obtains
\begin{equation} \label{eq:AF}
\frac{1}{8\pi\sqrt{s}}i\bar u_{2}\sum_{i=1}^{6}A_{i}M_{i}u_{1} = \chi_{2}^{\dag}\mathbf{F}\chi_{1} ,
\end{equation}
where the matrix $\mathbf{F}$ reads
\begin{align*}
\mathbf{F} &= i\mathbf{\sigma\cdot b}\,\mathcal{F}_{1}
 +  \mathbf{\sigma\cdot\hat q\,\sigma\cdot(\hat k\times b)}\mathcal{F}_{2} 
 + i\mathbf{\sigma\cdot\hat k\,\hat q\cdot b}\,\mathcal{F}_{3} \\
&\quad + i\mathbf{\sigma\cdot\hat q\,\hat q\cdot b}\,\mathcal{F}_{4}
 - i\mathbf{\sigma\cdot\hat q}\,b_{0}\mathcal{F}_{7}
 - i\mathbf{\sigma\cdot\hat k}\,b_{0}\mathcal{F}_{8} \,.
\end{align*}
Here, the hat over the three-vectors of course means normalization to a unit vector, 
and the four-vector $b_{\mu}$ is defined as
\begin{displaymath}
b_{\mu} = \epsilon_{\mu}-\frac{\mbox{\boldmath{$\epsilon$}}\cdot\mathbf{\hat k}}{|\mathbf{k}|}\,k_{\mu} \,.
\end{displaymath}
By substituting the standard representation of the Dirac spinors and matrices on the 
left-hand side of eq.~(\ref{eq:AF}), one finds the following expressions for the 
so-called CGLN-amplitudes $\mathcal{F}_{i}$:
\begin{align*}
\mathcal{F}_{1} &= (\sqrt{s}-m_{1})\frac{N_{1}N_{2}}{8\pi\sqrt{s}} \biggl[A_{1}
  +\frac{k\cdot q}{\sqrt{s}-m_{1}}A_{3} \\
& +(\sqrt{s}-m_{2}-\frac{k\cdot q}{\sqrt{s}-m_{1}})A_{4}-\frac{k^{2}}{\sqrt{s}-m_{1}}A_{6}\biggr] \,, \displaybreak[0]\\
\mathcal{F}_{2} &= (\sqrt{s}+m_{1})\frac{N_{1}N_{2}}{8\pi\sqrt{s}}
  \frac{|\mathbf{q}||\mathbf{k}|}{(E_{1}+m_{1})(E_{2}+m_{2})} \\
& \times\biggl[-A_{1}+\frac{k\cdot q}{\sqrt{s}+m_{1}}A_{3} \\ 
& + (\sqrt{s}+m_{2}-\frac{k\cdot q}{\sqrt{s}+m_{1}})A_{4}-\frac{k^{2}}{\sqrt{s}+m_{1}}A_{6}\biggr] \,, \displaybreak[0]\\
\mathcal{F}_{3} &= (\sqrt{s}+m_{1})\frac{N_{1}N_{2}}{8\pi\sqrt{s}}\frac{|\mathbf{q}||\mathbf{k}|}{E_{1}+m_{1}} \\
& \times \biggl[\frac{m_{1}^{2}-s+\frac{1}{2}k^{2}}{\sqrt{s}+m_{1}}A_{2}
  + A_{3}-A_{4}-\frac{k^{2}}{\sqrt{s}+m_{1}}A_{5}\biggr] \,, \displaybreak[0]\\
\mathcal{F}_{4} &= (\sqrt{s}-m_{1})\frac{N_{1}N_{2}}{8\pi\sqrt{s}}\frac{|\mathbf{q}|^{2}}{E_{2}+m_{2}} \\
& \times \biggl[\frac{s-m_{1}^{2}-\frac{1}{2}k^{2}}{\sqrt{s}-m_{1}}A_{2}
  + A_{3} - A_{4}+\frac{k^{2}}{\sqrt{s}-m_{1}}A_{5}\biggr] \,, \displaybreak[0]\\
\mathcal{F}_{7} &= \frac{N_{1}N_{2}}{8\pi\sqrt{s}}\frac{|\mathbf{q}|}{E_{2}+m_{2}}\biggl[(m_{1}-E_{1})A_{1} \\
& - \biggl(\frac{|\mathbf{k}|^{2}}{2k_{0}}(2k_{0}\sqrt{s}-3k\cdot q)\\
& \quad - \frac{\mathbf{q\cdot k}}{2k_{0}}(2s-2m_{1}^{2}-k^{2})\biggr)A_{2} \\
& + \bigl(q_{0}(\sqrt{s}-m_{1})-k\cdot q\bigr)A_{3} \\
& + \bigl(k\cdot q - q_{0}(\sqrt{s}-m_{1})+(E_{1}-m_{1})(\sqrt{s}+m_{2})\bigr)A_{4} \\
& + (q_{0}k^{2}-k_{0}k\cdot q)A_{5}-(E_{1}-m_{1})(\sqrt{s}+m_{1})A_{6}\biggr] \,, \displaybreak[0]\\
\mathcal{F}_{8} &= \frac{N_{1}N_{2}}{8\pi\sqrt{s}}\frac{|\mathbf{k}|}{E_{1}+m_{1}}\biggl[(E_{1}+m_{1})A_{1} \\
& + \biggl(\frac{|\mathbf{k}|^{2}}{2k_{0}}(2k_{0}\sqrt{s}-3k\cdot q)\\
& - \frac{\mathbf{q\cdot k}}{2k_{0}}(2s-2m_{1}^{2}-k^{2})\biggr)A_{2} \\
& + \bigl(q_{0}(\sqrt{s}+m_{1})-k\cdot q\bigr)A_{3} \\
& + \bigl(k\cdot q -q_{0}(\sqrt{s}+m_{1})+(E_{1}+m_{1})(\sqrt{s}-m_{2})\bigr)A_{4} \\
& - (q_{0}k^{2}-k_{0}k\cdot q)A_{5} - (E_{1}+m_{1})(\sqrt{s}-m_{1})A_{6}\biggr] \,,
\end{align*}
where
\begin{displaymath}
N_{i} = \sqrt{E_{i}+m_{i}}\,, \qquad E_{i} = \sqrt{\mathbf{p}_{i}^{2}+m_{i}^{2}}\,, \qquad i = 1,2 \,.
\end{displaymath}
We remark that, starting from eq.~(\ref{eq:TfiA}), we have utilized the Lorentz condition
$k\cdot\epsilon = 0$. This is also valid when electroproduction is considered, since 
in that case the object $\epsilon^{\mu}$ is proportional to the electron-photon vertex of QED, 
and the condition then follows from current conservation.

Restricting ourselves to s-and p-waves, we can use the CGLN-amplitudes to arrive at the 
multipoles $E_{0+}$, $M_{1+}$, etc.:
\begin{displaymath}
\left(\begin{array}{c} E_{0+}\\M_{1+}\\M_{1-}\\E_{1+}\end{array}\right) 
= \int_{-1}^{1} dz \!\left(\begin{array}{cccc}
  \frac{1}{2}P_{0} & -\frac{1}{2}P_{1} & 0                    & \frac{1}{6}P_{0,2} \\
  \frac{1}{4}P_{1} & -\frac{1}{4}P_{2} & -\frac{1}{12}P_{0,2} & 0 \\
 -\frac{1}{2}P_{1} &  \frac{1}{2}P_{0} &  \frac{1}{6}P_{0,2}  & 0 \\
  \frac{1}{4}P_{1} & -\frac{1}{4}P_{2} &  \frac{1}{12}P_{0,2} & \frac{1}{10}P_{1,3} \end{array}\right)
\!\!\left(\begin{array}{c} \mathcal{F}_{1}\\\mathcal{F}_{2}\\\mathcal{F}_{3}\\\mathcal{F}_{4}\end{array}\right)
\end{displaymath}
and
\begin{displaymath}
\left(\begin{array}{c} L_{0+}\\L_{1+}\\L_{1-}\end{array}\right) 
= \frac{k_{0}}{|\mathbf{k}|}\int_{-1}^{1} dz \left(\begin{array}{cc}
  \frac{1}{2}P_{1} & \frac{1}{2}P_{0} \\
  \frac{1}{4}P_{2} & \frac{1}{4}P_{1} \\
  \frac{1}{2}P_{0} & \frac{1}{2}P_{1} \end{array}\right)
\left(\begin{array}{c} \mathcal{F}_{7} \\ \mathcal{F}_{8} \end{array}\right) \,,
\end{displaymath}
where $P_{\ell}\equiv P_{\ell}(z)$ are the usual Legendre polynomials and $z$ is 
$\mathbf{\hat q}\cdot\mathbf{\hat k}$, {\it i.e.}\ the cosine of the scattering angle in the c.m.\ frame.
Furthermore, the abbreviations
\begin{align*}
P_{0,2} &= P_{0}-P_{2} &&\textrm{and} & 
P_{1,3} &= P_{1}-P_{3} 
\end{align*}
were used.

A word on units: Since we use 
\begin{displaymath}
\hbar = c = 1 \,, \qquad e^{2}=4\pi\alpha
\end{displaymath}
and normalize our Dirac spinors like $\bar u u = 2m $, the invariant amplitude $B_{5}$ 
has dimension $\textrm{GeV}^{-1}$, as can be seen, {\it e.g.}, from the contribution of the graph 
corresponding to the `Kroll-Ruderman` term. Therefore, the scattering amplitude $T_{fi}$ is 
dimensionless (see eq.~(\ref{eq:TfiB})), while the CGLN-amplitudes as well as the multipoles 
have dimension $\textrm{GeV}^{-1}$.

The unpolarized differential cross section for $\gamma p\rightarrow \mathrm{B}\,\mathrm{M}$ 
is given in terms of the CGLN-amplitudes as \cite{Donnachie}
\begin{align}
\frac{d\sigma}{d\Omega} &= \frac{|\mathbf{q}|}{|\mathbf{k}|}\bigg\{|\mathcal{F}_{1}|^{2}
  + |\mathcal{F}_{2}|^{2}+\frac{1}{2}|\mathcal{F}_{3}|^{2}+\frac{1}{2}|\mathcal{F}_{4}|^{2} \no \\
& + \Re(\mathcal{F}_{1}\mathcal{F}_{4}^{\ast}) + \Re(\mathcal{F}_{2}\mathcal{F}_{3}^{\ast}) \no \\
& + \bigl(\Re(\mathcal{F}_{3}\mathcal{F}_{4}^{\ast})-2\Re(\mathcal{F}_{1}\mathcal{F}_{2}^{\ast})\bigr)\cos\theta \no \\
& - \Bigl(\frac{1}{2}|\mathcal{F}_{3}|^{2}+\frac{1}{2}|\mathcal{F}_{4}|^{2}
  + \Re(\mathcal{F}_{1}\mathcal{F}_{4}^{\ast}+\mathcal{F}_{2}\mathcal{F}_{3}^{\ast})\Bigr)\cos^{2}\theta \no \\
& - \Re(\mathcal{F}_{3}\mathcal{F}_{4}^{\ast})\cos^{3}\theta \bigg\} \,.
\end{align}

\end{appendix}


\end{document}